\documentclass[twocolumn]{article}
\usepackage[left=2.5cm, right=2.5cm, top=2.5cm, bottom=2.5cm]{geometry}
\usepackage[utf8]{inputenc}
\usepackage[british]{babel}
\usepackage{authblk}
\usepackage{amsmath, amssymb}
\usepackage{graphicx}
\usepackage{xcolor}
\usepackage[colorinlistoftodos]{todonotes}
\usepackage{textcomp}
\usepackage{hyperref}
\usepackage[ruled,vlined]{algorithm2e}
\usepackage{bbm}
\usepackage{makecell}

\newcounter{comment}

\newcommand{\ABcolor}{violet} 
\newcommand{\JBcolor}{blue}
\newcommand{\HMcolor}{red}

{\refstepcounter{comment}%
\begin{quote}
\color{\ABcolor}\small$\blacksquare$ \textbf{\underline{Comment} $\sharp$\thecomment.~AB:}}%
{\end{quote}}

{\refstepcounter{comment}%
\begin{quote}
\color{\JBcolor}\small$\blacksquare$ \textbf{\underline{Comment} $\sharp$\thecomment.~JB:}}%
{\end{quote}}

{\refstepcounter{comment}%
\begin{quote}
\color{\HMcolor}\small$\blacksquare$ \textbf{\underline{Comment} $\sharp$\thecomment.~HM:}}%
{\end{quote}}

\DeclareMathOperator*{\argmin}{Arg\!min}
\DeclareMathOperator*{\argmax}{Arg\!max}

\DeclareMathOperator{\prox}{prox}
\newcommand{\inprod}[2]{\langle#1,#2\rangle}
\newcommand{\norm}[1]{\left\lVert#1\right\rVert}
\newcommand{\normBlock}[1]{{\left\vert\kern-0.25ex\left\vert\kern-0.25ex\left\vert #1 \right\vert\kern-0.25ex\right\vert\kern-0.25ex\right\vert}}

\providecommand{\keywords}[1]{\textbf{\textit{Keywords---}} #1}

\newcommand{\nonparametric}{non-parametric\xspace}
\newcommand{\quasimonochromatic}{quasi-monochromatic\xspace}
\newcommand{\quasiperiodic}{quasi-periodic\xspace}
\newcommand{\autocorrelation}{auto-correlation\xspace}

\newcommand{\eg}{\textit{e.g.}\xspace}
\newcommand{\ie}{\textit{i.e.}\xspace}

\title{Sparse data inpainting for the recovery of Galactic-binary gravitational wave signals from gapped data}
\author[1]{A.~Blelly\footnote{aurore.blelly@cea.fr}}
\author[1,2]{J.~Bobin}
\author[1]{H.~Moutarde}
\affil[1]{IRFU, CEA, Université Paris-Saclay, F-91191 Gif-sur-Yvette, France}
\affil[2]{IPARCOS, Complutense University, E-28040 Madrid, Spain}

\date{\today}

\begin{document}

\maketitle


\begin{abstract}
The forthcoming space-based gravitational wave observatory LISA will open a new window for the measurement of galactic binaries, which will deliver unprecedented information about these systems.
However, the detection of galactic binary gravitational wave signals is challenged by the presence of gaps in the data. 
Whether being planned or not, gapped data dramatically reduce our ability to detect faint signals and the risk of misdetection soars.
Inspired by advances in signal processing, we introduce a \nonparametric inpainting algorithm based on the sparse representation of the galactic binary signal in the Fourier domain. 
In contrast to traditional inpainting approaches, noise statistics are known theoretically on ungapped measurements only. This calls for the joint recovery of both the ungapped noise and the galactic binary signal. We thoroughly show that sparse inpainting yields an accurate estimation of the gravitational imprint of the galactic binaries. Additionally, we highlight that the proposed algorithm produces a statistically consistent ungapped noise estimate. We further evaluate the performances of the proposed inpainting methods to recover the gravitational wave signal on a simple example involving verification galactic binaries recently proposed in LISA data challenges.
\end{abstract}



\section{Introduction}
\label{sec:introduction}

\subsection{General Context}

As the first dedicated space-based gravitational wave observatory, LISA will considerably extend our experimental knowledge of gravitational waves. 
This device, composed of three satellites distant 2.5 million km one from another, will probe a frequency range so far unexplored. High hopes surround this mission as the scientific objectives are wide \cite{Audley:2017drz}: unique tests of general relativity, probes of galaxy constitution, studies of galactic binary population, estimation of stochastic background, detection of black hole mergers, multi-messenger astronomy \eg in conjunction with the ESA Athena mission \cite{Nandra:2013jka}, to name only few. 

The three satellites composing LISA will form the three arms of the interferometer. These arms will produce time series on 18 main data streams, which are then post-processed using time delay interferometry (TDI - denoted by the set $(X,Y,Z)$) in order to get rid of major noise sources such as laser noise. This remarkable procedure allows to retrieve a gravitational wave signal orders of magnitude smaller than the laser noise. However it becomes increasingly difficult to implement when adding realistic features to the description of the expected LISA signal, which motivates further conceptual developments \cite{Vallisneri:2020otf}. We will rather work here with the classical linear combination of the TDI denoted by $(A,E,T)$. This allows for a simplified noise representation since it is built to lead to a statistically uncorrelated noise between the three channels $(A,E,T)$ (see Ref.~\cite{Tinto:2014lxa} and references therein).

To retrieve gravitational wave signals from measurements, a large number of methods have been introduced, mainly in the framework of LIGO and Virgo. 
Nevertheless, they cannot apply directly to the future LISA data. 
For instance, and in contrast to LIGO and Virgo, LISA is expected to permanently record numerous signals emitted from various gravitational sources, among which Galactic Binaries (GBs) originating from white dwarfs, neutron stars or stellar-origin black holes and emitting \quasimonochromatic GWs.
Consequently one of the goals of the LISA mission is the resolution and characterization of several thousands of GBs \ie LISA will produce a wealth of information about GBs. Indeed 16 ultracompact binary systems --coined \emph{verification binaries} (VBs)-- have been electromagnetically identified so far \cite{Kupfer:2018}. Thanks to Gaia \cite{GaiaColl:2016} and LSST \cite{LSSTScienceColl:2009}, the detection of about 100 to 1000 VBs is foreseen \cite{Korol:2017qcx} before LISA starts observing. However 
the measurement of tens of thousands of white dwarf systems is expected during LISA lifetime (see Ref.~\cite{Lamberts:2019nyk} and references therein). 

Additionally, the LISA data will be distorted by various instrumental and observational effects such as imprints of glitches, noise with complex statistics or gaps. 
Accurately recovering information from these data mandates the development of new analysis techniques to address the associated data processing challenges. 
Further dealing with gapped data raises a significantly more complex challenge: interruptions of the measuring process can happen for many reasons, among which the telescope re-alignment, the transmission of data to Earth or simply the the occurrence of an unforeseen event on board. 
The duration and frequency of data gaps will vary depending on their origin.
On top of a manifest information loss, the presence of data gaps also induces an important information leakage which affects the identification process \cite{carre:2010arxiv, Edwards:2020tlp}. 
It will also induce noise correlations which were not initially present. Illustrations can be found in Sect.~\ref{subsec:prior_information}.

So far, considering the problem of source identification with a complex instrumental noise, many methods based on parameter estimation through maximum likelihood resolution (MCMC inference, matched filtering) were designed, especially for LISA mission \cite{crowder_solution_2007, littenberg_detection_2011, littenberg_global_2020}. 
To the best of our knowledge, the identification of GBs from gapped data has only been tackled with the Data Augmentation Algorithm \cite{baghi2019_DataAugmentation}. 
It has been developed for LISA data and aims at filling the data gaps with both signal and noise by trying to identify (through parameter estimation) the source that is present. 
To date, this algorithm was demonstrated for a single source only and not probed on cases with a potentially high unknown number of sources as expected with GBs in the LISA spectrum.

\paragraph*{Contributions :} In Ref.~\cite{blelly2020sparsity}, we introduced a novel, general, non-parametric framework for the detection and recovery of GBs. 
Built upon a sparsity-based modelling of the GBs, the proposed approach allows to account for the particular structures of TDI data as well as specificities of galactic binaries' waveform. 
Being non-parametric, it yields a fast low-bias estimate of the GBs signal, without prior knowledge of their number.
This new mathematical tool permits a precise detection of GBs, with an accurate control of the false discovery rate, which makes it an effective approach to robustly deal with the noise that contaminates the LISA data.

In the field of computer science, dealing with gapped data has long been considered in the framework of sparsity-based signal processing, leading to sparse data inpainting methods (see Refs. \cite{starck:book10, EladBook} and references therein). 
In this article, we therefore propose an innovative method that combines the non-parametric GBs recovery method we introduced in Ref.~\cite{blelly2020sparsity} and sparse data inpainting to mitigate the impact of data gaps on LISA science. 
Elaborating on our prior knowledge of the ungapped noise distribution, we estimate not only the missing signal but also the missing noise. 

The general context of our work, the corresponding framework and data modelling are presented in Section~\ref{sec:introduction}. 
Sections~\ref{sec:sparse_data_inpainting_LF} and \ref{sec:algorithms_and_implementation} describes the two algorithms that we developed to mitigate the impact of gaps on LISA data - detailed proofs and information can also be found in Appendix~\ref{sec:appendix}. 
Section~\ref{sec:experimental_results} focuses on assessing the performances of the two algorithms is various configurations. 
Last Section~\ref{sec:conclusion} draws conclusions and prospects over the present study.

\subsection{Notations}

Given two integers $N$ and $P$, consider a matrix $F \in \mathbbm{K}^{N \times P}$ with $\mathbbm{K} = \mathbbm{R} \text{ or } \mathbbm{C}$. 
It can be written as a series of rows $F_k$ or a series of columns $F_p'$:
\begin{equation}
F = \begin{bmatrix}
F_1 \\
F_2 \\
... \\
F_N
\end{bmatrix} = 
\begin{bmatrix}
F_1',F_2',..., F_P'
\end{bmatrix} \;,
\end{equation}
where $F_k \in \mathbb{K}^{P}, 1\leq k \leq N$ and $F_p' \in \mathbb{K}^{N},1\leq p \leq P$.
For a row $F_k$, we define the norm $s$ by:
\begin{equation}
\norm{F_k}_{s} = \left( \sum_{i=1}^P |F_k[i]|^s \right)^{1/s} \;.
\end{equation}
We can then define the norm $r,s$ on the matrix $F$ by:
\begin{equation}
\norm{F}_{r,s} = \left( \sum_{k=1}^N \norm{F_k}_s^r \right)^{1/r} \;.
\end{equation}

We write $A^*$ the conjugate transpose of the matrix $A$.
Finally, we will use the Hadamard product $\odot$ between a vector $\gamma \in \mathbbm{K}^N$ and a column vector $F_p'$, defined as:
\begin{equation}
\gamma \odot F_p' = \begin{bmatrix}
\gamma[1] F_p'[1] \\
\gamma[2] F_p'[2] \\
... \\
\gamma[N] F_p'[N]\\
\end{bmatrix} \;.
\end{equation}
The Hadamard product of a vector $\gamma$ and a matrix $F$ reads:
\begin{equation}
\gamma \odot F = \begin{bmatrix}
\gamma \odot F_1',\gamma \odot F_2',...,\gamma \odot F_P'
\end{bmatrix} \;.
\end{equation}

\subsection{Measurements modeling}

A gravitational event reaching LISA will generate a signal on each of the 3 Michelson-Morley-like arms of the detector. 
Three time series will then be produced by TDI \cite{Tinto:2014lxa}: the 3 signals $X$, $Y$ and $Z$ sampled at time $t_n = n \Delta T$ ($0 \leq n < N_T$). 
A linear combination of these TDI leads to the time series we will work with: the 3 data channels $A$, $E$ and $T$: 
\begin{eqnarray}
A[n] &=& \frac{Z[n] - X[n]}{\sqrt{2}} \;, \label{eq:def-channel-A} \\
E[n] &=& \frac{Z[n] - 2 Y[n] + X[n]}{\sqrt{6}} \;, \label{eq:def-channel-E} \\
T[n] &=& \frac{Z[n] + Y[n] + X[n]}{\sqrt{3}} \;, \label{eq:def-channel-T} 
\end{eqnarray}
whose noises are assumed to be statistically uncorrelated. 
Since the signal-to-noise ratio (SNR) of channel T is much smaller than those of channels A and E in the frequency range of interest, the mHz domain \cite{Prince:2002hp} --which is the relevant range for GB detection by LISA-- we will neglect $T$ and focus the signals from A and E \cite{Robson:2018jly}. 

We gather all the measurements in a single matrix, each column corresponding to a data channel:
\begin{equation}
\label{eq:def-measure-vector-U}
\mathcal{V} =\begin{bmatrix}
A[0] & E[0] \\
A[1] & E[1] \\
\vdots & \vdots \\
A[N_T-1] & E[N_T-1] \\
\end{bmatrix} \in \mathbbm{R}^{N_T \times 2} \;,
\end{equation} 
Channels $A$ and $E$ will be referred to as $\mathcal{V}_A$ and $\mathcal{V}_E$, and the measurements at time $t_n$ as $\mathcal{V}_A[n]$ and $\mathcal{V}_E[n]$. 
$\alpha$ will designate either channel $A$ or channel $E$. 
Thus $\mathcal{V}_\alpha$ denotes the content of the first or the second columns of the matrix of measurements.

There will be  interruptions in data taking, be it for planned device maintenance or unplanned reasons. 
In order to model these interruptions, we define the mask time function $m$ as:
\begin{equation}
m[n] = \begin{cases}
0 \quad \text{if the data is missing at time } t_n \\
1 \quad \text{otherwise}\\
\end{cases} \;.
\end{equation}

For each channel $\alpha \in \{A,E \}$, the measurement at time $n$ can be written as:
\begin{equation}
\label{eq:def-signal-noise-decomposition-time}
\mathcal{V}_\alpha[n] = m[n](\mathcal{S}_\alpha^{\text{true}}[n] + \mathcal{N}_\alpha^{\text{true}}[n]) \;,
\end{equation}
where $\mathcal{S}_\alpha^{\text{true}}[n]$ is the sought-for GW signal emitted by GB sources and received at time $t_n$, and $\mathcal{N}_\alpha^{\text{true}}[n]$ is the noise simultaneously measured.

It will later prove convenient to introduce a mask matrix:
\begin{align}
M &= \mathrm{diag}(\{m[n]\}_{0 \leq n \leq N_T-1}) \nonumber \\
&= \begin{bmatrix}m[0] & & \\ & \ddots & \\ & & m[N_T-1]\end{bmatrix} \;.
\end{align}
For any data in time domain $V \in \mathbbm{R}^{N_T \times 2}$, we will then denote the corresponding masked measurement by: 

\begin{equation}
M V = 
\begin{bmatrix}
m[0] V_A[0] & m[0] V_E[0] \\
m[1] V_A[1] & m[1] V_E[1] \\
\vdots & \vdots \\
m[N_T] V_A[N_T-1] & m[N_T] V_E[N_T-1] \\
\end{bmatrix}\;, 
\end{equation}
which is the matrix product between the mask $M$ and the measurement $V$.

The frequency domain is also discretized with a frequency step $\delta f$ depending on the number of measurements. 
Discrete frequencies correspond to the $f_k = k \delta f$, with $-N_f \leq k \leq N_f$ and $N_f = \lfloor N_T/2\rfloor$. 
Since we deal with real-valued signals, their negative ($-N_f \leq k \leq 0$) and positive ($0 \leq k \leq N_f$) frequency Fourier coefficients are complex-conjugated. 
In the following we will reserve the indices $n \in \{0, \ldots, N_T \}$ and $k \in \{-N_f, \ldots, N_f \}$ for the time and frequency variables respectively.

For a measurement $V_\alpha$ in channel $\alpha$, we define the  discrete Fourier transform: 
\begin{equation}
    \widetilde{V}_\alpha [k] = \frac{1}{\sqrt{N_T}} \sum_{n = 0}^{N_T -1} V_\alpha[n]    e^{-\frac{2 \i \pi k n}{N_T} } \;.
\end{equation}
We adopt the notation: 
\begin{equation}
\label{eq:def-measure-vector-U-hat}
\widetilde{V} = 
\begin{bmatrix}
\widetilde{V}_A[-N_f] & \widetilde{V}_E[-N_f] \\
\vdots & \vdots \\
\widetilde{V}_A[+N_f] & \widetilde{V}_E[+N_f] \\
\end{bmatrix}\in \mathbbm{C}^{(2N_f+1) \times 2} \;, 
\end{equation}
for the signals in the Fourier domain similarly to our time-domain convention (\ref{eq:def-measure-vector-U}). 

Considering a matrix $M \in \mathbbm{R}^{n\times n}$, we will use the following algebra notations:

\begin{description}
\item[$\mathrm{Ker}(M)$:] kernel of $M$ (vectors $U$ of $\mathbbm{R}^n$ such that $M U = 0$),
\item[$\mathrm{Ran}(M)$:] range of $M$ (image of $\mathbbm{R}^n$ by $M$),
\item[$I$:] identity matrix of $\mathbbm{R}^{n\times n}$.
\end{description}

In App.~\ref{appendix:frequently_used_variables} we provide a synthesis of our writing conventions with the list of the most frequently used variables of this study.
In particular, calligraphic letters are used for data, estimators (denoted by a hat) and the sequential solutions of optimization problems (denoted by sequence index $n$).
Capital letters are used for dummy variables.

\subsection{Modeling of gaps, noise and signal}

\label{subsec:prior_information}
The specificity of the problem of LISA gapped data is that we have prior information on both the signal and the noise in {\it absence of gaps}. 
We present these priors in the following paragraphs and explain how the gaps impact the priors. 

\subsubsection{Different types of data gaps}
Several types of gaps are likely to happen:
\begin{description}
\item[Long, unplanned gaps:] During LISA Pathfinder mission, a 5-day-long data gap occurred, during which the device went into safety mode and could not collect data. This can also happen during the LISA Mission : the device will not always be able to collect data. The \emph{mission duty cycle} is estimated around $75\%$, which means that there will be around $25\%$ of missing data in the form of long duration gaps. 
\item[Maintenance gaps] Regularly the antennas need to be realigned. This process is expected to take place once every two weeks and last approximately 7 hours.
\item[Short, unplanned gaps] It is still possible to have daily short events that will cause data-taking interruption. Such events are likely to happen every day, over a short time period (from a few seconds to a few hours). For instance, this could be the result of "gapping" the signal in presence of data glitches. 
\end{description}
We will show below the impact of the three different types of gaps on the signal.

\subsubsection{Prior on noise distribution}
We will consider an additive ungapped noise with the following properties: 

\begin{enumerate}
    \item Gaussianity: in time domain and frequency domain, the noise obeys a gaussian law $\mathcal{G}(m,\Gamma)$ of mean $m$ and variance $\Gamma$. 
    \item Stationnarity: the noise \autocorrelation function is left invariant by time translations.
    \item Zero-mean: in Fourier domain, the noise has a null mean value \cite{LDCwebsite}.
\end{enumerate}

Under these hypotheses, the distribution followed by the noise on each channel $\alpha \in \{A,E\}$ in Fourier space explicitly reads:
\begin{equation}
\label{eq:FD_noise_distribution}
\widetilde{\mathcal{N}}_\alpha^{\text{true}}[k] = N_{1,\alpha}[k] + i N_{2,\alpha}[k]  \;, 
\end{equation}
where $N_{1,\alpha}[k], N_{2,\alpha}[k]$ are random variables following a standard normal (\ie reduced centered) law:
\begin{equation}
N_{1,\alpha}[k] \sim \mathcal{G}(0,\mathbf{\Sigma}_\alpha), \quad N_{2,\alpha}[k] \sim \mathcal{G}(0,\mathbf{\Sigma}_\alpha) \;.
\end{equation}
with $\mathbf{\Sigma}_\alpha[k]$ the noise power spectral density (PSD), \ie the Fourier transform of the \autocorrelation function. 

Last but not least, for a measurement $V \in \mathbb{R}^{N_T \times 2}$ we will often use the corresponding whitened data in frequency domain:
\begin{equation}
\mathbf{\Sigma}^{-1/2} \widetilde{V} =
\begin{bmatrix}
\mathbf{\Sigma}^{-1/2}_A \widetilde{V}_A & \mathbf{\Sigma}^{-1/2}_E \widetilde{V}_E
\end{bmatrix}\;.
\end{equation}

\begin{figure}[h!]
    \centering
    \includegraphics[width=0.5\textwidth]{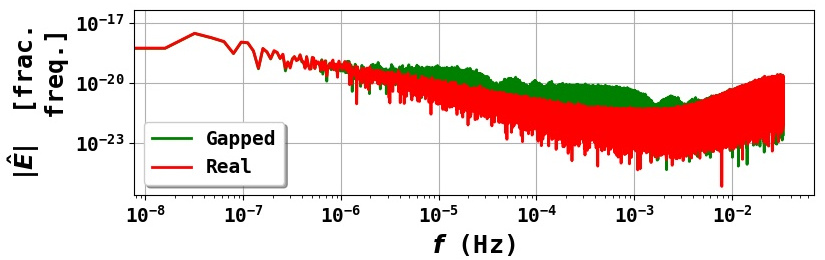} 
    \caption{Fourier transform of simulated noise for an ungapped signal (in red) and a gapped signal (in green) for daily short unplanned gaps and maintenance gaps (see simulation details in Sec. \ref{subsec:prior_information},\ref{sec:experimental_results}). The presence of gaps in time domain leads to a modification of the expected noise distribution in Fourier domain. This modification impacts the mHz frequency band of interest. }
    \label{fig:impact_gaps_noise_distribution}
\end{figure}

Fig. \ref{fig:impact_gaps_noise_distribution} represents the noise amplitude in frequency domain for the ungapped signal (labelled as "real") and the gapped signal (labelled as "gapped").
The presence of gaps created a distortion in the noise PSD as well as added correlation between the different frequencies (whereas the noise is initially supposed not to contain any correlation between frequencies in Fourier domain).


\section{Sparse Data Inpainting}
\label{sec:sparse_data_inpainting_LF}

In this section, we introduce a sparsity-constrained inpainting algorithm. Following Ref.~\cite{blelly2020sparsity}, we will primarily focus on GW signals coming from GBs.

\subsection{Sparse signal modelling and recovery}

\begin{figure}[h!]
    \centering
\begin{tabular}{c}
\includegraphics[width=0.45\textwidth]{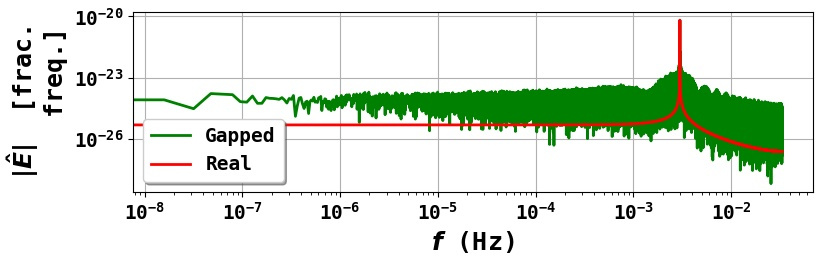} \\
\includegraphics[width=0.45\textwidth]{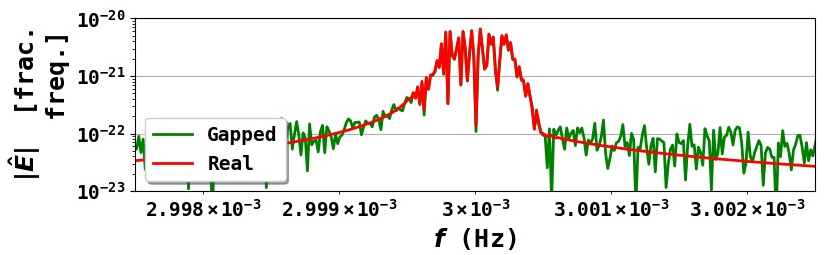}
\end{tabular}
    \caption{Simulated signal received from a GB in frequency domain: in red the real, ungapped signal, and in green the gapped signal (small and planned gaps). 
    The gaps in time domain lead to a deformation of the signal in frequency domain: the gapped signal is "less sparse" than the ungapped signal. The more information we lose, the greater the deformation of the signal. 
    }
    \label{fig:impact_gaps_waveform}
\end{figure}

The temporal gravitational signature of GBs are \quasimonochromatic signals on the whole mission duration. More precisely, the expected waveform received by LISA is known analytically, and has been well studied so far \cite{blaut_mock_2010}. GB signatures can be parameterized with 8 variables, one of them being the GB frequency $f_0$ around which the signal is emitted. We chose to start with these sources as they present a close-to-sinusoidal morphology, which leads to a very simple representation in the Fourier domain, as shown by the signal labelled as "real" in Fig.\ref{fig:impact_gaps_waveform}. As we emphasized in Ref.~\cite{blelly2020sparsity}, being nearly mono-frequency the GBs waveform admits a naturally sparse distribution in the harmonic domain (Fourier basis). In other words, it can be approximated accurately with a small number of Fourier components.

In Ref.~\cite{blelly2020sparsity} we specifically focused on the recovery of GB signals from noisy GW data. In this context, we developed a sparsity-enforcing denoising algorithm derived from extensively used methods in the signal and image processing community. These methods are based on the hypothesis that there exists a representation domain in which the information related to the sought signal is concentrated into few coefficients whereas noise is not. In such a sparse domain, the SNR is maximized for the sought signal \cite{Bruckstein2009, starck:book10} which is the key for building an efficient denoising procedure. 

This framework is particularly well adapted for GBs since they have a well-known waveform, which is nearly monochromatic in Fourier domain. At the same time, the noise is assumed to be Gaussian in the same domain, with a known power spectral density. These assumptions allowed us to develop a method  \cite{blelly2020sparsity} that can separate the total signal coming from GBs (as the sum of all signals coming from GBs) from the noise. The gist of such a method amounts to estimating the signal by looking for it as the minimizer of the following cost function:
\begin{align}
\widehat{\mathcal{S}} = \argmin_{\widetilde{S} \in \mathbbm{C}^{(2N_f+1)\text{x}2}} &\Bigg[ \underbrace{\norm{\gamma \odot   \mathbf{\Sigma}^{-1/2} \widetilde{S}}_{1,2}}_{\text{Sparsity constraint}} \nonumber \\ 
&+ \underbrace{\frac{1}{2}  \norm{\widetilde{\mathcal{V}}-\widetilde{S}}_{2,2, \mathbf{\Sigma}}^2}_{\text{Data fidelity}}  \Bigg]
\;,
\label{eq:general_ungapped-problem_formulation}
\end{align}
where:
\begin{equation}
\norm{\widetilde{V}}_{2,2, \mathbf{\Sigma}}^2 = \norm{\mathbf{\Sigma}^{-1/2}\widetilde{V}}_{2,2}^2 \;.
\end{equation}
This cost function is composed of two terms. The first one aims at constraining the solution to be sparse in the Fourier domain. The second is a data fidelity term that favors the solution to be as close as possible to the observed data $\mathcal{Y}$, with regard to the expected noise distribution in the frequency domain. As showed in Ref.~\cite{blelly2020sparsity}, the optimization problem Eq.~(\ref{eq:general_ungapped-problem_formulation}) admits a unique minimizer that can be computed analytically:
\begin{equation}
\label{eq:L12_solution}
\widetilde{\widehat{\mathcal{S}}}_\alpha[k] = 
\begin{cases}
\displaystyle \frac{\norm{\mathbf{\Sigma}^{-1/2}\widetilde{\mathcal{V}}[k]}_{1,2} - \gamma[k]}{\norm{\mathbf{\Sigma}^{-1/2}\widetilde{\mathcal{V}}[k]}_{1,2}} \widetilde{\mathcal{V}}_\alpha[k] \\ \qquad \text{if } \norm{\mathbf{\Sigma}^{-1/2}\widetilde{\mathcal{V}}[k]}_{1,2} > \gamma[k] \\
0 \\ \qquad  \text{otherwise}
\end{cases} \;.
\end{equation}
Using a $\chi^2_{4}$-law with 4 degrees of freedom, $\gamma$ is a threshold that is set with regard to noise level:
\begin{equation}
\label{eq:x0-from-p-value}
\mathbbm{P}(\chi^2_{4} \geq \gamma^2) = \rho \;,
\end{equation}
where $\rho$ is the chosen rejection rate for the hypothesis test $H_0$: "there is no GW signal at frequency $k$" against $H_1$: "here is a GW signal at frequency $k$" (see Ref.~\cite{blelly2020sparsity} for a detailed discussion).

The proposed sparsity-based denoising method has been proved to yield a low-bias estimation of a large number of GB signals in a non-parametric way. Quite interestingly, this doesn't require either their identification nor the prior knowledge of their number. Furthermore, such a method allows for a robust detection procedure, with a guaranteed control of the false positive rate. This makes it well adapted to deal with the LISA noise.


\subsection{Estimating sparse signals from gapped data}
\label{sec:estimating_signal_with_gapped_data}

In this section, we detail how we build upon the denoising method introduced in Ref.~\cite{blelly2020sparsity} to further deal with gapped data.

\begin{figure}[h!]
    \centering
\begin{tabular}{c}
\includegraphics[width=0.4\textwidth]{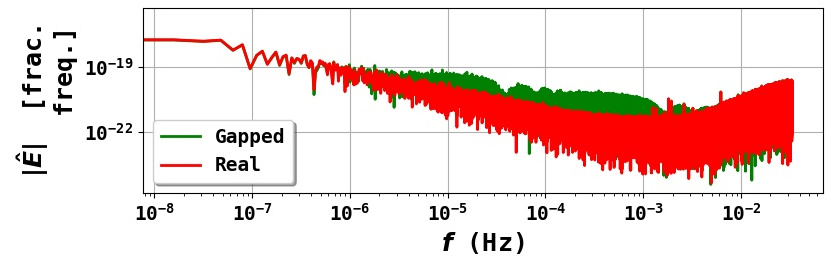} \\
\includegraphics[width=0.4\textwidth]{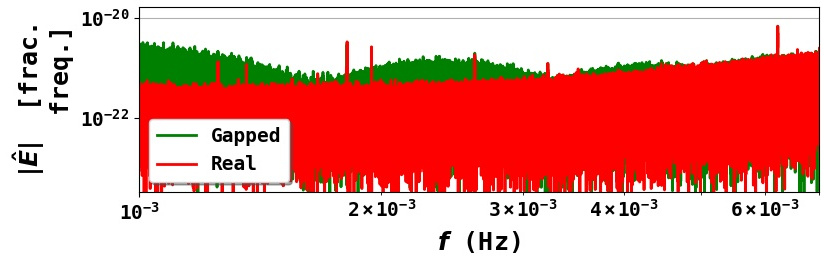} 
\end{tabular}
    \caption{Fourier transform of LDC1-3 data \cite{LDCwebsite}, composed of 10 verification binaries, for the real ungapped signal (red) and the signal that we gapped with daily short unplanned gaps and maintenance gaps (green).
    Signals that could have been identified without gaps are now completely drowned in noise.}
    \label{fig:information_leakage}
\end{figure}

The presence of gaps in the data leads to a significantly more challenging problem:
\begin{description}
\item[Loss of signal power:]  In presence of gaps, the waveform is distorted and is not sparse anymore: Fig.~\ref{fig:impact_gaps_waveform} shows the waveform resulting from a gapped signal in the Fourier domain (labelled as "gapped"). Even if the main part of the waveform seems unchanged (Fig.~\ref{fig:impact_gaps_waveform}, lower row), there is a gap-dependent power loss that is due to the information originating from the incompleteness of the data. In practice, this can lead to imprecise detection and identification.
\item[Noise leakage:] Fig.~\ref{fig:information_leakage} shows how gaps impact both the signal and the noise. The red plot features the Fourier transform of data without gaps, whereas the green plot corresponds to the Fourier transform of the gapped data. 
This experiment illustrates how gaps also result in a noise leakage, and radically change the statistics of the noise in the Fourier domain. With respect to noise, gaps again hamper the detection and identification of GBs as a significantly larger amount of GB signals will be drowned in noise in the case of a gapped signal.
\end{description}

In order to mitigate the impacts of gaps on the data, we developed two algorithms that enable to fill the gaps for both the signal and the noise without requiring the identification of the underlying individual sources.

From the viewpoint of signal processing, gapped data are more generally equivalent to partiallly observed or incomplete data. In such a case, the model described in Eq.~(\ref{eq:def-signal-noise-decomposition-time}) admits an infinite number of solutions. Fortunately, based on the sparsity property of the sought-after GB signals in the Fourier domain, this problem finds a solution in the framework of compressed sensing (CS) \cite{candes_sparsity_2007,donoho_compressed_2006}. In a nutshell, CS is a signal sampling/recovery theory that builds upon two pillars: 

\begin{description}
    \item[Sparsity:] the first ingredient of compressed sensing is the sparsity of the signal to be recovered in a known signal representation. In the present article, this assumption naturally applies to the GB signals in the Fourier domain.
    \item[Incoherence:] the second pillar of compressed sensing is the incoherence between the domain where the samples are taken and the one in which the signal to be retrieved is assumed to be sparse. In brief, the incoherence between the two domains implies that if a given signal is sparse in the Fourier domain, it is in contrast smeared out in the sample domain, where data are missing. Mathematically speaking, incoherence can also be interpreted as a generalized Heisenberg uncertainty principle \cite{Candes06Uncertainty}.
\end{description}

From the very large literature related to CS in applied mathematics, it has been showed that the signal can be retrieved with high accuracy by solving a regularized least-squares optimization problem (see Ref.~\cite{EldarGitta_Book12} and references therein), even more with severely incomplete data. 
More specific to signal and image processing, the special case of signal recovery from gapped data is better known as \emph{inpainting} \cite{starck:jalal06,starck:book10} .

In the present article, applying CS to the recovery of GB signals from gapped data would boil down to minimizing the following cost function:
\begin{align}
\widetilde{\widehat{\mathcal{S}}} = \argmin_{\widetilde{S} \in \mathbbm{C}^{(2N_f+1)\text{x}2}} &\Bigg[ \norm{\gamma \odot   \mathbf{\Sigma}^{-1/2}_{gap} \widetilde{S}}_{1,2}\nonumber \\ 
&+ \frac{1}{2}  \norm{\widetilde{\mathcal{V}}-M\widetilde{S}}_{2,2, \mathbf{\Sigma}_{gap}}^2  \Bigg]
\;,
\label{eq:general_gapped-problem_formulation_0}
\end{align}
where $\mathbf{\Sigma}_{gap}$ is the noise PSD of the gapped noise, which we do not know \textit{a priori}. It could be computed based on the expected noise distribution and the mask, but the computation would be costly and the corresponding problem would still remain hard to solve.

Standard CS methods generally make the assumption that the noise statistics is known at the level of the measured data. 
However, this is not the case for the LISA data: one generally has access to some knowledge about the theoretical PSD of the {\it ungapped} noise in the Fourier domain. 

Accounting for the right noise distribution is particularly important as it allows one to carefully control the false positive rate of the sought-after GB signal \cite{blelly2020sparsity}. 
This is however a key discrepancy with respect to standard methods since the theoretical noise statistics has to be modelled accurately in the Fourier domain. 
We therefore propose an innovative sparse inpainting algorithm that retrieves both the ungapped signal and noise.

To that end, we first define a new estimator, $\widehat{\mathcal{U}}$, of the missing data within the gaps (both signal and noise).
We have $\widehat{\mathcal{U}} \in \mathrm{Ran}(I-M) = \mathrm{Ker}(M)$.

Now, we can rewrite the problem as a joint estimation of the signal and the missing data, in a form that is really similar to that of Eq.~(\ref{eq:general_ungapped-problem_formulation}):

\begin{eqnarray}
(\widehat{\mathcal{S}},\widehat{\mathcal{U}})  =  \argmin_{\substack{S,U \in \mathbbm{R}^{N_T\text{x}2} \\ M U = 0 }} \Bigg[ & & \norm{\gamma \odot   \mathbf{\Sigma}^{-1/2} \widetilde{S}}_{1,2} \label{eq:general_gapped-problem_formulation} \\ 
& + & \frac{1}{2}  \norm{\widetilde{\mathcal{V}} + \widetilde{U}- \widetilde{S}}_{2,2, \mathbf{\Sigma}}^2  \Bigg] 
\;, \nonumber 
\end{eqnarray}
where $\widehat{\mathcal{S}}$ stands for the estimator of total ("ungapped") GW signal and $\widehat{\mathcal{U}}$ the estimator of the missing data within the gaps. 
This way, both the priors on the signal and the noise are still valid, which allows the known theoretical PSD of the "ungapped" noise to be used.

Let us notice that the resulting optimization problem now exhibits a mixed formulation with terms expressed both in time and frequency. Consequently, it does not admit a closed-form solution and the minimizer needs to be computed numerically with an iterative algorithm, which is described in the next section.

\section{Algorithms \& Implementation}
\label{sec:algorithms_and_implementation}

In the scope of LISA data processing, the goal of data inpainting is twofold. On the one hand, and following the approach introduced in Ref.~\cite{blelly2020sparsity}, it aims at providing an estimation of the total signal originating from GBs that is robust with respect to noise and gaps. On the second hand, data inpainting can be deemed a general pre-processing step for LISA data treatment, whose objective is to deliver estimated ungapped data. The latter more specifically emphasize on an accurate signal estimation with no power loss as well as an estimate of the noise that matches the statistics of the expected one. Doing so, the resulting inpainted data can be used as inputs to classical identification techniques such as bayesian inference. In that regard the sparsity framework has the advantage of being a \nonparametric methodology, which works independently from the actual number of GW sources. To that purpose we hereafter introduce two algorithms that tackle these two views of the inpainting problem.

\subsection{Resolution Algorithm}

The problem in Eq.~(\ref{eq:general_gapped-problem_formulation}) benefits from certain properties, which can be used to build an efficient minimization algorithm. Indeed, it is not only a strictly convex problem, but also a block-convex problem: it is convex with regard to the variables $V$ while $U$ is kept fixed and vice versa. Let us now denote the global cost function to be minimized as:
\begin{equation}
J(S,U) = \norm{\gamma \odot   \mathbf{\Sigma}^{-1/2} \widetilde{S}}_{1,2} + \frac{1}{2}  \norm{\widetilde{\mathcal{V}} + \widetilde{U} - \widetilde{S}}_{2,2, \mathbf{\Sigma}}^2 \;.
\label{eq:cost-function_gapped-problem}
\end{equation}
Thanks to the block-convexity of the problem in Eq.~(\ref{eq:general_gapped-problem_formulation}) both variables $S$ and $U$ can be sequentially and iteratively updated. For that purpose, we make use of a Block Coordinate Descent (BCD) algorithm \cite{Xu2013_BCD} which can be summarized with the two following steps:
\begin{align}
\begin{cases}
\mathcal{U}^{n+1} &= \displaystyle \argmin_{\substack{U \in \mathbbm{R}^{N_T\text{x}2} \\ M U = 0 }} J(\mathcal{S}^n,U)  \nonumber \\
\mathcal{S}^{n+1} &= \displaystyle \argmin_{S\in \mathbbm{R}^{N_T\text{x}2}} J(S,\mathcal{U}^{n+1}) \\
\end{cases} \;,
\label{eq:BCD_algorithm}
\end{align}
with initialization $\mathcal{S}^{0} = 0$. The sequence $(\mathcal{S}^n,\mathcal{U}^n)$ converges to the sought estimators $(\widehat{\mathcal{S}},\widehat{\mathcal{U}})$, the solution of Eq.~(\ref{eq:general_gapped-problem_formulation}) \cite{Tseng:2001}. We now detail each of these steps. 

\subsubsection*{Updating the noise}
Let us introduce the estimator of the ungapped noise $\widehat{\mathcal{N}}$. It is related to the other estimators by the following equation:

\begin{equation}
\underbrace{\mathcal{V} + \widehat{\mathcal{U}}}_{\text{Completed data}} = \widehat{\mathcal{S}} +  \widehat{\mathcal{N}} \;.
\label{eq:estimators_relation}
\end{equation}

For the sake of simplicity, we can rewrite the proposed BCD-based algorithm so that the global signal and the noise {\it in the gaps} are computed sequentially. To that end, let us consider the update of $\mathcal{U}^{n+1}$: 
\begin{align}
\mathcal{U}^{n+1} &= \argmin_{\substack{U \in \mathbbm{R}^{N_T\text{x}2} \\ M U = 0 }} J(\mathcal{S}^n, U) \nonumber \\
&= \argmin_{\substack{U \in \mathbbm{R}^{N_T\text{x}2} \\ M U = 0 }} \frac{1}{2} \norm{\widetilde{\mathcal{V}} + \widetilde{U} - \widetilde{\mathcal{S}}^n}^2_{2,2,\mathbf{\Sigma}} \;.
\label{eq:missing_data_equation}
\end{align}
Introducing $N$ the noise variable defined by the following change of variables:
\begin{equation}
N = \underbrace{\mathcal{V}+ U}_{\text{Observed noisy signal}} - \underbrace{\mathcal{S}^n}_{\text{estimated noiseless signal}}\;,
\label{eq:change_of_variable}
\end{equation}
one can recast Eq.~(\ref{eq:missing_data_equation}) as an equation on noise.
We define $\mathcal{N}^n_{\text{gap}} = \mathcal{V} - M \mathcal{S}^n$ as the estimated noise outside of the gaps at iteration $n$.
This eventually leads to the following equivalent problem:
\begin{equation}
\mathcal{N}^{n+1} = \argmin_{\substack{ N \in \mathbbm{R}^{N_T\text{x}2}  \\ \\ \mathcal{N}^n_{\text{gap}} = M N }} \frac{1}{2} \norm{\widetilde{N}}^2_{2,2,\mathbf{\Sigma}} \;,
\label{eq:classical-inpainting_noise-update}
\end{equation}
where $\mathcal{N}^{n}$ converges to the noise estimator $\widehat{\mathcal{N}}$ for $n\to +\infty$

Solving this problem is challenging because the noise PSD $\mathbf{\Sigma}$ is known in frequency domain while the equality constraint $\mathcal{N}^n_{\text{gap}} = M N$ is defined in time domain. This problem does not admit a closed-form expression. 
To evaluate numerically its minimizer, we use the Chambolle and Pock primal-dual algorithm algorithm \cite{chambolle:hal-00490826}. 
This algorithm has two main advantages: i) it remains computationally simple, and ii) it can further be preconditioned to speed up convergence, which is particularly convenient since the noise PSD is ill-conditioned. 
We refer to Appendix~\ref{app:noise_solver_CP-algorithm} for more details about this algorithm.

\subsubsection*{Updating the signal}

From Eq.~(\ref{eq:classical-inpainting_noise-update}), we get that:
\begin{align}
\mathcal{U}^n &= (I-M) \mathcal{U}^n \\
&= (I-M)(\mathcal{S}^n + \mathcal{N}^{n+1} - \mathcal{V}) \\
&= (I-M)(\mathcal{S}^n+\mathcal{N}^{n+1}) \;.
\end{align}
We can then introduce the updated data $\mathcal{V}^n$  as
\begin{align}
\mathcal{V}^n &= \mathcal{V} + \mathcal{U}^n \\
&= \mathcal{V} + (I-M)(\mathcal{S}^n + \mathcal{N}^{n+1}) \;,
\end{align}
which is the data whose gaps have been filled in at the $n$-th iteration. 
Consequently, the signal estimation step can be recast as the estimation of the signal over the ungapped data $\mathcal{V}^{n+1}$:
\begin{align}
\mathcal{S}^{n+1} = \argmin_{S \in \mathbbm{R}^{N_T\text{x}2}} &\Bigg[ \norm{\gamma \odot   \mathbf{\Sigma}^{-1/2} \widetilde{S}}_{1,2} \nonumber \\
&+ \frac{1}{2}  \norm{\widetilde{\mathcal{V}}^{n+1}-\widetilde{S}}_{2,2, \mathbf{\Sigma}}^2  \Bigg] \;.
\end{align}
Quite remarkably, thanks to the particular structure of our inpainting algorithm, the update of the signal $\mathcal{S}^{n+1}$ can be done from {\it pseudo measurements} where the noise is inpainted. 
This allows to use the theoretical PSD rather than the more complex PSD of the gapped noise. The resulting update is then equivalent to the denoising problem investigated in Ref.~\cite{blelly2020sparsity}. 
More precisely, the signal $\mathcal{S}^{n+1}$ admits a closed-form expression which takes the form of a specific thresholding applied to $\widetilde{\mathcal{V}}^{n+1}$ in the Fourier domain. 
It was further highlighted that a block-sparsity regularization leads to significantly better detection and estimation precision. 
This results in a block-based thresholding, which we adopt in the following. 
More details about this denoising procedure are provided in Ref.~\cite{blelly2020sparsity}.

The overall algorithm of Classical Inpainting (C.I.) is described in Alg.~\ref{alg:algorithm_gapped-problem_lower-frequencies}. 

\begin{algorithm}[h!]
 \KwIn{$\mathcal{V},M,\mathbf{\Sigma},\epsilon$}
 \textbf{Initialization:}  $\mathcal{S}^0 = 0 $\;
 \While{$\max |\mathbf{\Sigma}^{-1/2}(\widetilde{\mathcal{S}}^{n+1}-\widetilde{\mathcal{S}}^n)| > \epsilon$}{
    $\mathcal{N}_{\text{gap}}^{n+1} = \mathcal{V} - M \mathcal{S}^{n}$\;
     $\mathcal{N}^{n+1} = \displaystyle\argmin_{\mathcal{N}_{\text{gap}}^{n+1} = M N} \frac{1}{2} \norm{\widehat{N}}^2_{2,2,\mathbf{\Sigma}}$ \;
      $\mathcal{V}^{n+1} = \mathcal{V} + (I-M)(\mathcal{S}^{n}+\mathcal{N}^{n+1})$ \;
      $
    \begin{aligned}
          \mathcal{S}^{n+1} = \displaystyle \argmin_{S} &\norm{\gamma \odot \mathbf{\Sigma}^{-1/2} \widetilde{S}}_{1,2} \\
          &+ \norm{\widetilde{\mathcal{V}}^{n+1}-\widetilde{S}}^2_{2,2,\mathbf{\Sigma}}\;
    \end{aligned}
    $
 }
 \caption{Classical Data Inpainting Algorithm (C.I.)}
 \label{alg:algorithm_gapped-problem_lower-frequencies}
\end{algorithm}

\subsection{Behavior of the inpainted noise in a gap}
\label{sec:behavior_BF-IPT}

\begin{figure}[h!]
    \centering
    \begin{tabular}{c}
\includegraphics[width=0.5\textwidth]{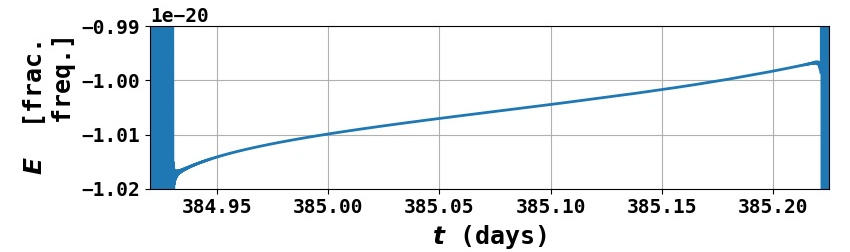}\\
\includegraphics[width=0.48\textwidth]{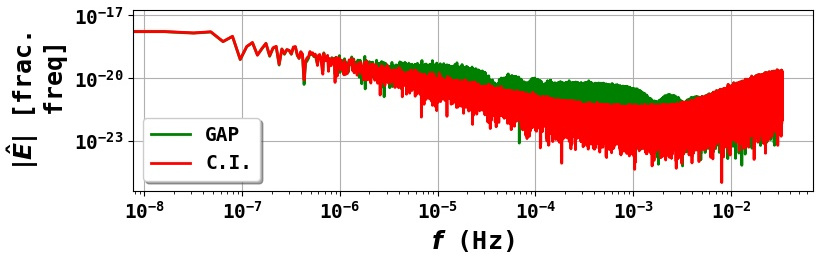}
    \end{tabular}
    
    \caption{\textbf{Upper row:} Inpainted noise in time domain in a gap for classical inpainting algorithm (\textbf{C.I.}). We only recover noise frequencies for which the correlation length is superior to the gap size. 
    \textbf{Lower row:} Comparison between gapped signal and inpainted signal in frequency domain: the noise spectrum was flattened according to the expected distribution.}
    \label{fig:BF_inpainting_TD-solution-behavior_in-gap}
\end{figure}

In this section, we illustrate the performances of the proposed inpainting algorithm especially on typical LISA noise. 
For that purpose, we consider simulated noise realizations for LISA (not containing any gravitational signal), whose statistics is described by the theoretical PSD of the LDC1-3. 
Gapped data are generated with small unplanned gaps and planned gaps (see Section \ref{sec:experimental_results} for more details about gaps generation). 
Next, inpainting is applied to the resulting noise according to Eq.~(\ref{eq:classical-inpainting_noise-update}). 
The solution is displayed in Fig.~\ref{fig:BF_inpainting_TD-solution-behavior_in-gap} with a zoom on a single gap in the time domain (upper row); the Fourier transform of the global solution is also represented (lower row). 

In the time domain, the inpainted noise inside a gap only exhibits a low-frequency smooth morphology, which already allows for a good signal extraction. 
Indeed, this algorithm is completely deterministic and cannot {\it generate} information lost in the gaps. 
More specifically, such an inpainting methods basically fills out gaps by exploiting correlations that go beyond the size of the gaps. 
This entails that within a gap, the retrieved information content tends to have low frequencies, corresponding to wavelengths larger than the size of the gap. 
Higher frequency information cannot be recovered and is definitely lost.  
As we will see more precisely below, this results in an \emph{unavoidable power loss} on the frequency spectrum of the recovered noise compared to the expected distribution. 
Even if in Fig.~\ref{fig:BF_inpainting_TD-solution-behavior_in-gap} the inpainted noise seems to follow the expected noise PSD. 
This shows that Eq.~(\ref{eq:classical-inpainting_noise-update}) acts like a low-pass filter with regard to the ungapped data. 

In order to provide a more consistent noise inpainting procedure, we describe in the following section an extension of our algorithm.


\subsection{Modified Sparse Data Inpainting}
\label{sec:sparse_data_inpainting_allF}
\label{sec:modified_noise_IPT_strategy}

Let us recall that the objective of our inpainting approach is twofold: designing an inpainting algorithm that i) provide an efficient detection and reconstruction of the GB signal and ii) more generally yield a statistically consistent inpainted noise. In the previous section, we pointed out that a traditional sparsity-enforced inpainting does not reach the second objective as it leads to a gap-dependent noise power loss. In this paragraph, we propose extending sparse inpainting to further correct for this effect.

A straightforward approach would consist in adding a high-frequency stochastic term to this low-frequency solution within the gaps in time domain. However, this would only produce a poor solution, as the added high-frequency noise in the gaps would not be compatible with the one observed outside of the gaps. For instance, it would not be continuous at the boundaries of the gaps, which would likely create high frequency artifacts. The main challenge then boils down to finding a high frequency noise correction that matches both the expected distribution and the boundary condition on the border of every gap. For that purpose, we propose now a method that combines the approach of Alg.~\ref{alg:algorithm_gapped-problem_lower-frequencies}  with the use of a stochastic term in order to create an appropriate inpainted noise solution. 

\paragraph{Generation of a compatible high frequency term:} Let us draw an ungapped random sample $\mathcal{N}_{\text{samp}}$ that follows the expected noise distribution. To that end, we define the function $f_{LF}(V)$ as:
\begin{equation}
f_{LF}(V) = \argmin_{\substack{N \in \mathbbm{R}^{N_T\text{x}2} \\ \\ M N = V }} \frac{1}{2} \norm{\widetilde{N}}^2_{2,2,\mathbf{\Sigma}} \;.
\label{eq:BF-algo_equation}
\end{equation}
Let $D_N$ be the difference:
\begin{equation}
D_N = f_{LF}(M\mathcal{N}_{\text{samp}})-\mathcal{N}_{\text{samp}} \;,
\label{eq:HF_noise_component}
\end{equation}
leading to a signal term, whose value is $0$ out of the gaps (since $f_{LF}(M\mathcal{N}_{\text{samp}})$ is exactly $\mathcal{N}_{\text{samp}}$ out of the gaps). Additionally, it only contains information at high frequency since the low-frequency content in $f_{LF}(M\mathcal{N}_{\text{samp}})$ has been removed (see Section~\ref{sec:behavior_BF-IPT}).
Owing to Eq.~(\ref{eq:HF_noise_component}) the difference $D_N$ is null outside the gaps. Consequently, it can be virtually added to any gapped measurement that has been inpainted with C.I. without altering the noise PSD. 

The resulting modified noise inpainting writes:
\begin{align}
N &= \underbrace{f_{LF}(\mathcal{V})}_{\text{Low frequency}}+\underbrace{f_{LF}(M \mathcal{N}_{\text{samp}}) - \mathcal{N}_{\text{samp}}}_{\text{High frequency}} \nonumber \\
&= f_{LF}(U+M \mathcal{N}_{\text{samp}}) - \mathcal{N}_{\text{samp}} \;,
\label{eq:noise_all-frequencies_solution}
\end{align}
where Eq.~(\ref{eq:BF-algo_equation}) exhibits a \emph{linearity property with regard to its input}, as demonstrated in Appendix~\ref{app:linearity_CP-algo}.

\paragraph{Modified Inpainting (M.I.) Algorithm:} The modified algorithm builds upon the C.I. algorithm by integrating sequentially in the global process the signal extraction step. We further prove in Appendix~\ref{app:all-frequencies_convergence} that the overall process consists in minimizing a cost function that shares similarities with the standard inpainting defined in Eq.~(\ref{eq:cost-function_gapped-problem}), but to which we added a correcting term. It can be solved using the exact same minimization scheme based on the BCD architecture up to a modification of the input and output. More precisely, the resulting algorithm is detailed in  Alg.~\ref{alg:all-frequencies_inpainting_algorithm}.

\begin{algorithm}[h!]
 \KwIn{$\mathcal{V},M,\mathbf{\Sigma},\epsilon$}
 \textbf{Initialization:}  $\mathcal{S}^0 = 0$ , draw $\mathcal{N}_{\text{samp}}$\;
  \While{$\max |\mathbf{\Sigma}^{-1/2}(\widetilde{\mathcal{S}}^{n+1}-\widetilde{\mathcal{S}}^n)| > \epsilon$}{
    $\mathcal{N}_{\text{gap}}^{n+1} = \mathcal{V} - M \mathcal{S}^{n} + M \mathcal{N}_{\text{samp}}$\;
     $\mathcal{N}^{n+1}_{LF} = \displaystyle\argmin_{\mathcal{N}_{\text{gap}}^{n+1} = M N} \frac{1}{2} \norm{\widetilde{N}}^2_{2,2,\mathbf{\Sigma}}$ \;
     $\mathcal{N}^{n+1} = \mathcal{N}^{n+1}_{LF}-  \mathcal{N}_{\text{samp}}$ \;
      $\mathcal{V}^{n+1} = \mathcal{V} + (I-M)(\mathcal{S}^{n}+\mathcal{N}^{n+1}) $ \;
      $
    \begin{aligned}
          \mathcal{S}^{n+1} = \displaystyle \argmin_{S} &\norm{\gamma \odot \mathbf{\Sigma}^{-1/2} \widetilde{S}}_{1,2} \\
          &+ \norm{\widetilde{\mathcal{V}}^{n+1}-\widetilde{S}}^2_{2,2,\mathbf{\Sigma}}\;
    \end{aligned}
    $
 }
 \caption{Modified Data Inpainting Algorithm (M.I.)}
 \label{alg:all-frequencies_inpainting_algorithm}
\end{algorithm}

\begin{figure}[h!]
    \centering
    \begin{tabular}{c}
\includegraphics[width=0.45\textwidth]{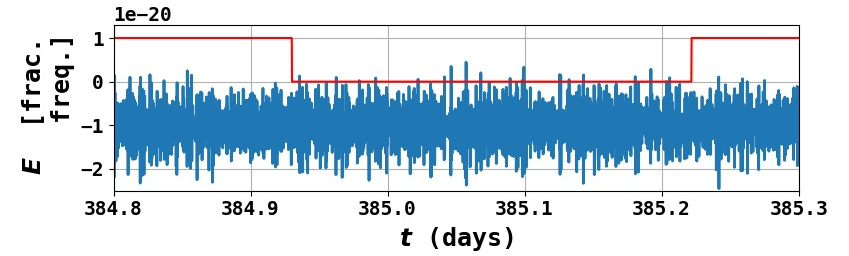}\\
\includegraphics[width=0.43\textwidth]{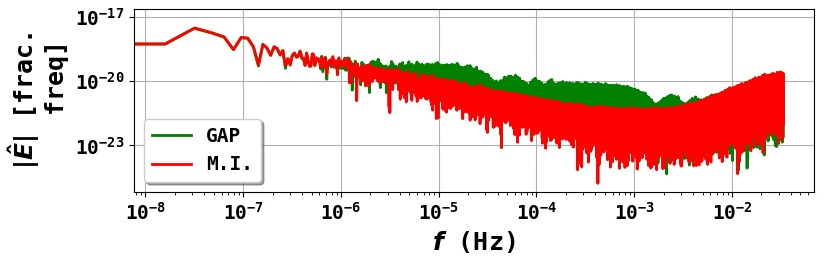}
    \end{tabular}
    
    \caption{\textbf{Upper row:} Inpainted noise in time domain in a gap for modified inpainting algorithm (\textbf{M.I.}). The red line shows the position of the gap.
    \textbf{Lower row:} Comparison between gapped signal and inpainted signal in frequency domain: the noise spectrum was flattened according to the expected distribution .}
    \label{fig:allF_inpainting_TD-solution-behavior_in-gap}
\end{figure}

To further evaluate the impact of the new algorithm on the inpainted noise, we make the exact same test as in Section~\ref{sec:behavior_BF-IPT}.
Fig.~\ref{fig:allF_inpainting_TD-solution-behavior_in-gap} shows the solution which has been inpainted with the proposed algorithm. 
This time the real signal and the inpainted one do not visually show any differences.

The power loss was corrected using the added stochastic term. Performances of both algorithms are assessed in details in Section~\ref{sec:experimental_results}.


\subsection{Implementation}

\subsubsection{Compensating the power loss}

Although Alg.~\ref{alg:algorithm_gapped-problem_lower-frequencies} smoothens the effective PSD, the final estimated noise PSD does not match the theoretical one (expected distribution) as showed by Fig.~\ref{fig:median_D-KL}.
Indeed the estimated PSD (after noise inpainting) and the expected one are identical up to a multiplicative coefficient that empirically matches with the amount of data that was lost. 
This phenomenon is called \emph{power loss}, and is mainly related to the fact that the inpainting algorithm \textbf{C.I.} can only inpaint lower frequencies - this can be easily understood by looking at Fig.~\ref{fig:BF_inpainting_TD-solution-behavior_in-gap}.

This power loss impacts the detection capacity of the \textbf{C.I.} algorithm since we set the threshold $\gamma$, as reminded in  Section~\ref{sec:estimating_signal_with_gapped_data}, with regard to the \emph{theoretical} noise distribution whereas it should be set with regard to the \emph{effective} distribution. Since the effective PSD is lower than the expected one due to power loss, it corresponds to choosing a threshold that is higher than what it should be, therefore increasing the risk of non-detection. We correct it by adjusting the noise level, computing an effective PSD $\mathbf{\Sigma}_{eff}$ as follow:
\begin{equation}
    \mathbf{\Sigma}_{eff} = r_{data} \mathbf{\Sigma} \;,
\end{equation}
with :
\begin{equation}
    r_{data} = \frac{\displaystyle\sum_{n} m[n]}{N_T} \;,
\end{equation}
the proportion of remaining information. This first-order correction yields a decent description of the noise distribution after the use of the \textbf{C.I.} algorithm, but becomes imperfect as the proportion of lost data increases. 

\subsubsection{Improving the speed of convergence}

From the viewpoint of optimization, the more data is missing the slower the convergence of the iterative minimization algorithms is. Limiting this phenomenon can be done by making use of the Fixed Point Continuation or FPC \cite{Yin08b}, which advocates computing sequential estimates with decreasing regularization parameters. In the present context, instead of setting a fixed threshold $\gamma$ as explained in Ref.~\cite{blelly2020sparsity}, it is first set to a large value and then decreased towards the final objective threshold $\gamma_{final}$. Implementing the FPC significantly improves the convergence speed.


\section{Experimental Results}

\label{sec:experimental_results}

The performances of our algorithms are assessed on three main outcomes: (i) the detection capacity (comparatively to the ungapped case), (ii) the statistics of the noise and (iii) the quality of extracted signal. 

The result of the two first outcomes are totally independent from the method used to extract the signal. As mentioned above, this work could be used as a pre-processing step in the LISA pipeline, and we evaluate the performances of the algorithms with this application in mind. We also assess the impact of gaps by estimating the quality of the extracted signal compared to the signal we would have extracted without gaps. This part is entirely dependent on the chosen extraction method.
Finally the proposed algorithms are evaluated on the realistic simulations LDC1-3 \cite{LDCwebsite} to which gaps have been added. 

\subsection{Gaps generation} 

Data gaps are characterized by two parameters: their duration $L_{gap}$ and their period $T_{gap}$ (time period over which we observe one gap). In Section~\ref{subsec:prior_information} we described the three different types of gaps that we consider here. The numerical values used for this study are reported in Table~\ref{tab:gaps_parameters}.

\begin{table}[h!]
    \centering
    \begin{tabular}{|c|c|c|}
    \hline
Type & Duration $L_{gap}$  & Period $T_{gap}$ \\
\hline
Small & 10 min  & Every $24$ hours\\
\hline
Medium & 7 hours  & Every 2 weeks \\
\hline
Large & 3 days  & Every 12 days\\
\hline
    \end{tabular}
    \caption{Description of the three types of gaps used for the study. Small and large gaps are both unplanned, whereas medium sized gaps represent the planned interruptions (or maintenance gaps). We simulate one gap of length $L_{gap}$ once every $T_{gap}$.}
    \label{tab:gaps_parameters}
\end{table}

For small and medium-sized gaps, we split the signal in consecutive blocks of length $T_{gap}$. For each block, we uniformly randomly position a gap of length $L_{gap}$. Additionally two consecutive large gaps of length $L_{gap} = 3 \text{ days}$ are separated by a duration given by a Poisson law\footnote{Private communication from N. Korsakova.} of parameter $\lambda = 9$.

\subsection{Noise inpainting evaluation}

\subsubsection{Independence of the solution with respect to the chosen sample}

\begin{figure}[h!]
    \centering
\begin{tabular}{c}
\includegraphics[height=0.2\textheight]{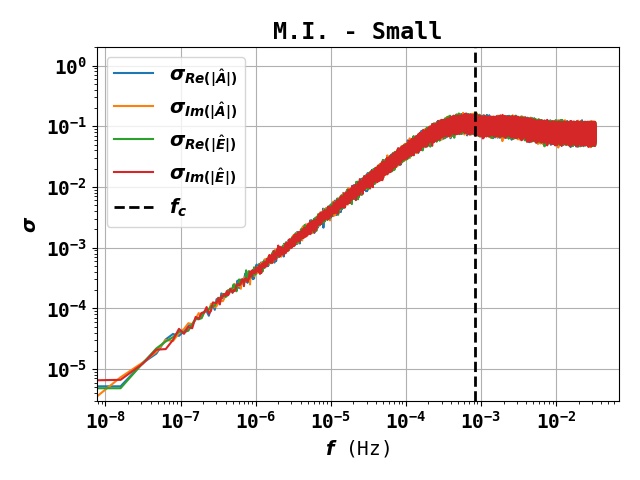}\\
\includegraphics[height=0.2\textheight]{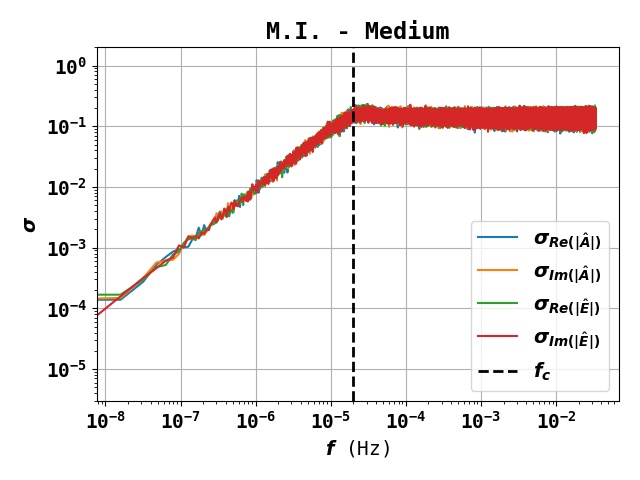}  \\
\includegraphics[height=0.2\textheight]{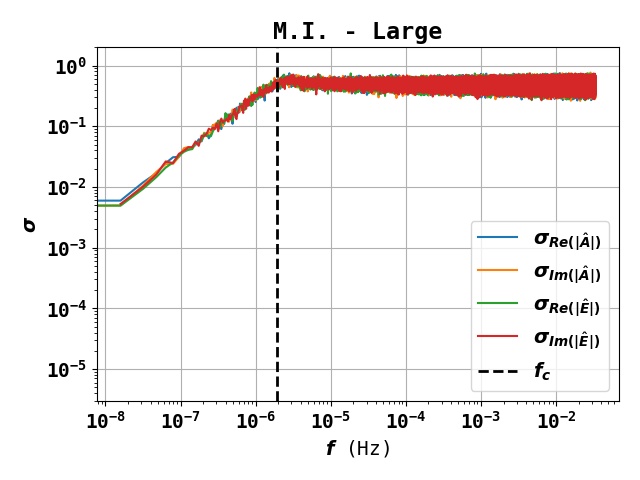}
\end{tabular}
    \caption{Distribution over 50 samples $\mathcal{N}_{\text{samp}}$ of real and imaginary parts of 
    $\mathbf{\Sigma}^{-1/2}\left( \widetilde{f_{LF}}(M(\mathcal{N}^{true}+\mathcal{N}_{\text{samp}})) - \widetilde{\mathcal{N}}_{sample} - \widetilde{\mathcal{N}}^{true} \right)$
    , with $\mathcal{N}^{true}$ the input noise (before gaps) and $f_{LF}(M(\mathcal{N}^{true}+\mathcal{N}_{\text{samp}})) - \widetilde{\mathcal{N}}_{samp}$ the recovered inpainted noise for M.I. algorithm and sample $\mathcal{N}_{\text{samp}}$.
    \textbf{Upper row:} Small gaps. 
    \textbf{Middle row:} Medium gaps. 
    \textbf{Lower row:} Large gaps.  
    The algorithm acts like a low-pass filter of order 1. The sample that we use in order to fill the gaps only impacts frequencies higher than the cut-off frequency $f_c$ (represented as a dashed black line), which is directly linked to the gap size. The cut-off frequency depends on the gap duration $L_{gap}$ (in seconds) as: $f_c = 1/(2 L_{gap})$.
    }
    \label{fig:sampling_independance-of-solution}
\end{figure}

Since the M.I. algorithm Alg.~\ref{alg:all-frequencies_inpainting_algorithm} makes use of an extra stochastic term, we first evaluate its sensitivity to the particular drawing of $\mathcal{N}_{\text{samp}}$. 
The answer is partially contained in Alg.~\ref{alg:all-frequencies_inpainting_algorithm} itself. 
We explained in Section~\ref{sec:modified_noise_IPT_strategy} that the lower-frequency part of the solution only depends on the measurements --meaning that it is independent from $\mathcal{N}_{\text{samp}}$-- whereas the higher frequency component mainly depends on $\mathcal{N}_{\text{samp}}$. 
In order to assess this dependence, let us consider a single input of the form:
\begin{equation}
\mathcal{V}_\alpha = M \mathcal{N}_\alpha \;.
\end{equation}
with $N_\alpha$ a noise sample following the expected ungapped noise distribution.
Inpainted noise solution are then computed with the M.I. algorithm Alg.~\ref{alg:all-frequencies_inpainting_algorithm} for various draws of $\mathcal{N}_{\text{samp}}$.

Fig.~\ref{fig:sampling_independance-of-solution} displays the standard deviation of the difference between the real noise and the solutions obtained for each sample in Fourier domain, frequency by frequency, for different types of gaps. 
This difference has further been whitened with the inverse theoretical noise PSD. Fig.~\ref{fig:sampling_independance-of-solution} quantitatively shows that inpainting leads to a low-pass filter effect confirming the qualitative features discussed in Sections~\ref{sec:behavior_BF-IPT} and \ref{sec:modified_noise_IPT_strategy}). 
Additionally, we point out two more features of the algorithms: the presence of a cut-off frequency and the behavior of the maximum deviation to the real noise distribution. 

The cut-off frequency, labelled as $f_c$ on the plot and defined as:
\begin{equation}
f_c = \frac{1}{2 L_{gap}}\;,
\label{eq:cut-off_frequency}
\end{equation}
matches the effective cut-off frequency of the filter-like behavior of Alg.~\ref{alg:all-frequencies_inpainting_algorithm}. 
As the gap duration increases, the cut-off frequency decreases: the noise component can only be rightfully recovered when its half-wavelength is superior to the size of the gap. 
Below the cut-off frequency, the recovered noise is quite close to the real noise. 
Above this frequency, there is a deviation that becomes more important as the amount of lost data increases.

\subsubsection{Recovered noise distribution}
\label{subsec:divergence_kullback-leibler}

\begin{figure}[h!]
    \centering
\begin{tabular}{c}
\includegraphics[height=0.2\textheight]{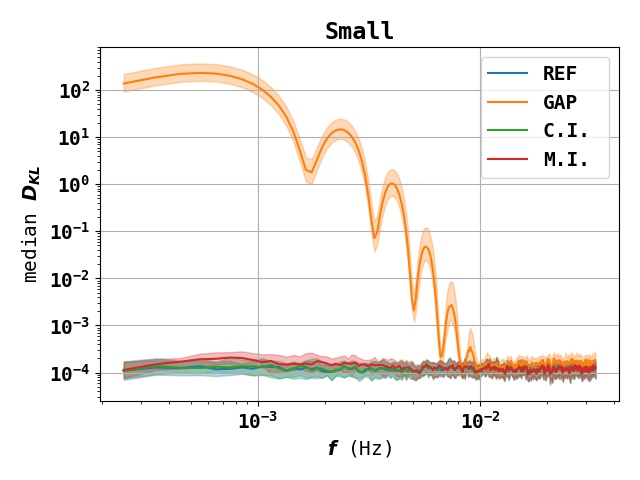} \\
\includegraphics[height=0.2\textheight]{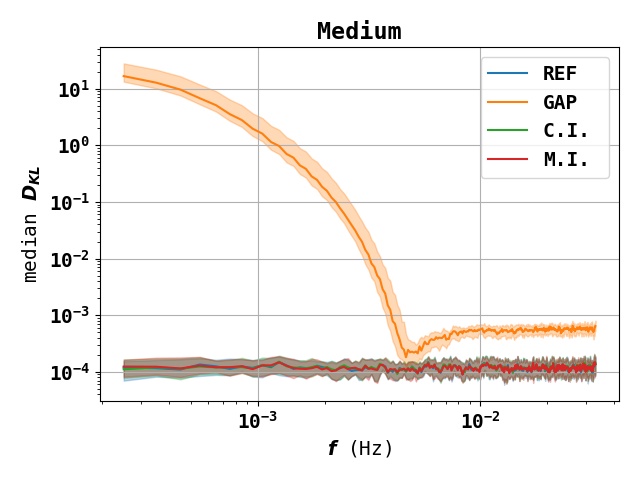}  \\
\includegraphics[height=0.2\textheight]{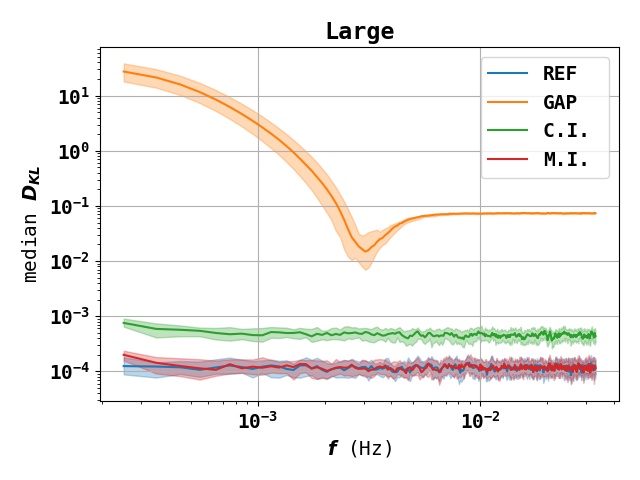}
\end{tabular}
    \caption{    Evolution of Kullback-Leibler divergence with frequency for reference ungapped signal (blue), gapped signal (orange), signal inpainted with C.I. (green) and signal inpainted with M.I. (red) for different types of gaps.
    \textbf{Upper row:} Small gaps. \textbf{Middle row:} Medium gaps. \textbf{Lower row:} Large gaps. 
    Power loss becomes more important as the gap size increases, which explains the incompatibility between reference noise statistic and C.I. noise statistic. 
    Gaps impact noise distribution in the mHz band in a non-negligible way. C.I. is effective for small- and medium-sized gaps, but less on large gaps as it cannot make up for the power loss. }
    \label{fig:median_D-KL}
\end{figure}

We emphasized previously that Alg.~\ref{alg:algorithm_gapped-problem_lower-frequencies} and Alg.~\ref{alg:all-frequencies_inpainting_algorithm} could be used as a pre-processing step for filling out gaps prior to performing further analysis such as event identification. 
To that end, we assess the impact of the inpainting on the noise statistics using the Kullback-Leibler divergence as a performance indicator. 

\paragraph{Kullback-Leibler divergence ($D_{KL}$):} The Kullback-Leibler divergence \cite{kullback1951} measures a discrepancy between the expected and the recovered noise distributions. 
We normalize the noise estimator $\widehat{\mathcal{N}}$ (limit of the sequence $(\mathcal{N}^{n})_{n\in\mathbbm{N}}$) in frequency domain as:
:
\begin{equation}
\eta_\alpha = \mathbf{\Sigma}^{-1/2} \widetilde{\widehat{\mathcal{N}}}_\alpha \;.
\end{equation}
If the recovered noise follows the expected law, we expect both the real part and the imaginary part of $\eta_\alpha$ to be drawn from a reduced centered normal law at any frequency $f$. 
Thus, we will measure the discrepancy of the law of the normalized noise compared to the reduced centered normal law. 

Under the approximation that the recovered noise follows a Gaussian law at any frequency $f$, the Kullback-Leibler divergence writes \cite{cover-book:2012}:
\begin{equation}
D_{KL}(f) = \frac{1}{2} \big(  \sigma(f)^2 + \mu(f)^2 - 1 - \ln{\sigma(f)^2} \big) \;,
\end{equation}
where $\sigma(f)$ is the measured standard deviation at frequency $f$ and $\mu(f)$ is the estimated expectation for the same frequency. 

As we only have one noise realization at each frequency $f$, we estimate the expectation and the standard deviation over a small neighborhood of frequencies around $f$. 
For a frequency $f_k$, let us define a neighborhood of size $n_k$:
\begin{equation}
I_{n_k}(f_k) = \{ f_j \displaystyle \}_{k-\frac{n_k}{2} \leq j \leq k+\frac{n_k}{2}} \;.
\end{equation}
Then defining:
\begin{equation}
\Omega(f_k) = \{ Re(\eta_\alpha^j), Im(\eta_\alpha^j), j \in I_{n_k}(f_k), \alpha \in \{A,E \} \} \;,
\end{equation}
an approximation of Kullback-Leibler divergence is given by:
\begin{equation}
\widetilde{D}_{KL}(f_k) = \frac{1}{2} \Big(  \mathbbm{V}[\Omega(f_k)] + \mathbbm{E}[\Omega(f_k)]^2 - 1 - \ln{\mathbbm{V}[\Omega(f_k)]} \Big) \;,
\end{equation}
where $\mathbbm{V}[\Omega(f_k)]$ and $\mathbbm{E}[\Omega(f_k)]$ respectively denote the variance and the expectation over the elements of $\Omega(f_k)$.

For the plots, we chose a window width of $\Delta f = 0.1~\text{ mHz}$ with an overlap of frequencies between two consecutive estimations of the divergence.

\paragraph{Experiment:}

For an input constituted of noise only, as:
\begin{equation}
\mathcal{V}_\alpha = M \mathcal{N}_\alpha \;,
\end{equation}
we computed the solutions given by the two  inpainting algorithms for various input noises $N_\alpha$ and various samples $\mathcal{N}_{\text{samp}}$. 
We compared the final noise distribution to the expected distribution through the Kullback-Leibler divergence. 

Fig.~\ref{fig:median_D-KL} represents the KL divergence evolution with frequency over 50 samples $(M, \mathcal{N}_\alpha, \mathcal{N}_{\text{samp}})$, when different types of gaps are present, for the ungapped signal, the gapped signal and the inpainted signals.
Looking at the discrepancy of the gapped signal, we note that the frequency band of interest (the mHz band) is impacted by the presence of gaps, whatever the type of gaps. 
Small but frequent gaps impact most the expected noise distribution (top plot). 
However, these are also the easiest type of gaps to deal with, as C.I. is enough to correct the noise distribution in Fourier domain. 
Planned gaps (middle plot), even though they are not that wide, show the limits of this inpainting algorithm. 
As the amount of data loss becomes more important, it cannot make up for the power loss: this explains the inconsistency between the expected distribution (blue) and the inpainted distribution (green). The M.I. algorithm can handle this power loss (red). 
Large gaps (bottom plot) associated with important amount of data loss, are the most difficult to mitigate as the corresponding power loss is even more sensible than for planned gaps. 
This case makes plain the necessity to use M.I. in order to compensate this huge power loss. 

To summarize: the two algorithms help mitigate the impact of data gaps. Alg.~\ref{alg:algorithm_gapped-problem_lower-frequencies} cannot compensate the power loss, but reaches a noise distribution that is --up to a multiplicative coefficient-- similar to the expected noise distribution. 
However scrutinizing the local noise statistic in the neighborhood of a gap, one finds noise statistics that are very different to the expected distribution.
On the contrary Alg.~\ref{alg:all-frequencies_inpainting_algorithm} enables to reach final noise statistics matching the expected statistics both locally and globally, correcting the power loss along the way. 

\subsection{Impact on detection capacity}
\label{subsec:FPR}

The following experiment evaluates the global performance of the overall algorithm (combining inpainting and signal extraction) in terms of false positive detection rate, \ie in proportion of signals that are detected where there is no corresponding input signal.

\begin{figure}[h!]
    \centering
\begin{tabular}{c}
\includegraphics[height=0.2\textheight]{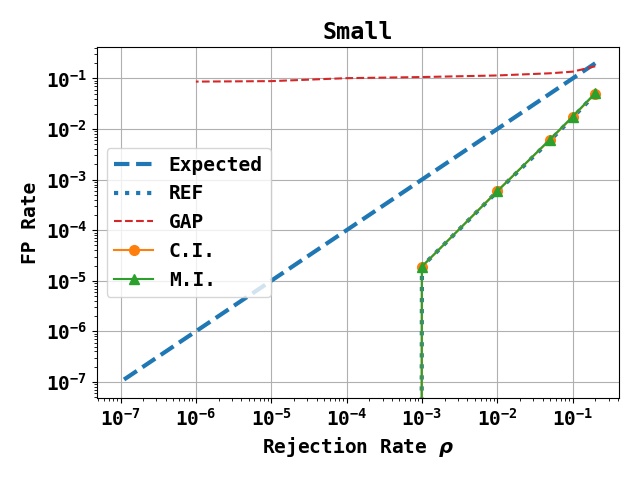}\\
\includegraphics[height=0.2\textheight]{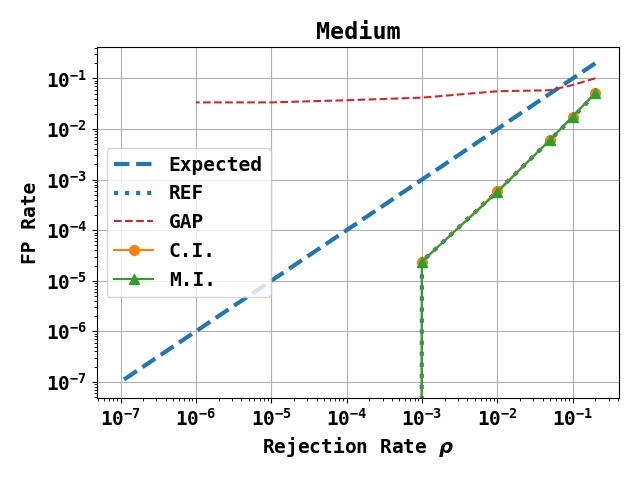}  \\
\includegraphics[height=0.2\textheight]{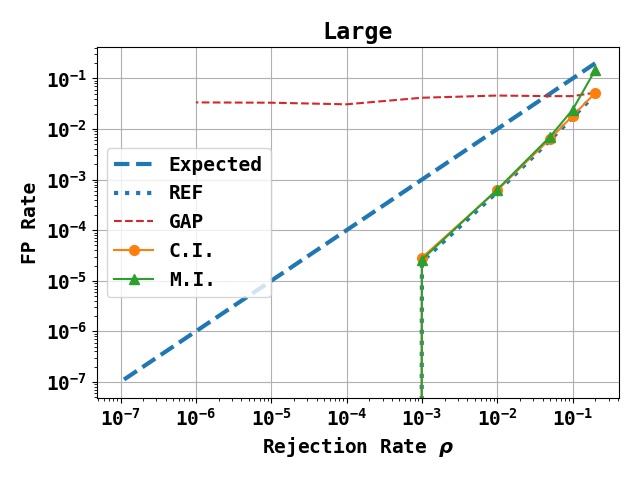}
\end{tabular}
    \caption{FPR with rejection rate for real ungapped signal (blue), gapped signal (red), signal inpainted with C.I. (orange) and signal inpainted with M.I. (green) for different types of gaps. \textbf{Upper row:} Small gaps. \textbf{Middle row:} Medium gaps. \textbf{Lower row:} Large gaps. 
    The FPR for algorithms C.I. and M.I. is identical to the one obtained for an ungapped signal.}
    \label{fig:median_FPR}
\end{figure}

In order to assess the detection capacity, we evaluated the number of false positive (FP) signals. For a given rejection rate $\rho$ (defined as in Section~\ref{sec:sparse_data_inpainting_LF}), we estimate the FP rate as:
\begin{equation}
    R_{FP}(\rho) = \frac{\text{\#FP}}{N_f} \;,
    \label{eq:def_FP_rate}
\end{equation}
where \#FP denotes the number of frequency detected as signal at the end of the algorithms for an initial input composed of noise only.

Considering an input constituted of noise only as:
\begin{equation}
\mathcal{V}_\alpha = M \mathcal{N}_\alpha \;,
\end{equation}
and a fixed mask $M$, we computed the solutions given by the two inpainting algorithms for various input noises $\mathcal{N}_\alpha$ and various samples $\mathcal{N}_{\text{samp}}$. 
We then computed the FP rate corresponding to these experiments for various rejection rate $\rho$. 
The results are presented in Fig.~\ref{fig:median_FPR}. 
For all types of gaps, the FP rate obtained for both inpainting algorithms is identical to the one obtained for ungapped data. 
Thus the combination of signal extraction and noise inpainting does not create any kind of detection artifact as the FP rate after inpainting is identical to the one estimated for an ungapped signal.

\subsection{Quality of the extracted signal}

We now assess the performances of the overall algorithms (combining both inpainting and signal extraction) in terms of quality of recovered signal. 
To that end, computing the normalized mean square error (NMSE) between the expected signal and the recovered signal quantifies the loss that originates from the gaps and which cannot be recovered.

\begin{figure}[h!]
    \centering
\begin{tabular}{c}
\includegraphics[height=0.2\textheight]{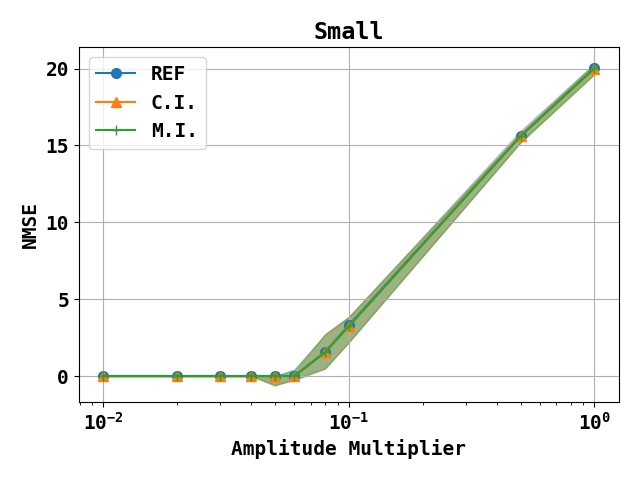} \\
\includegraphics[height=0.2\textheight]{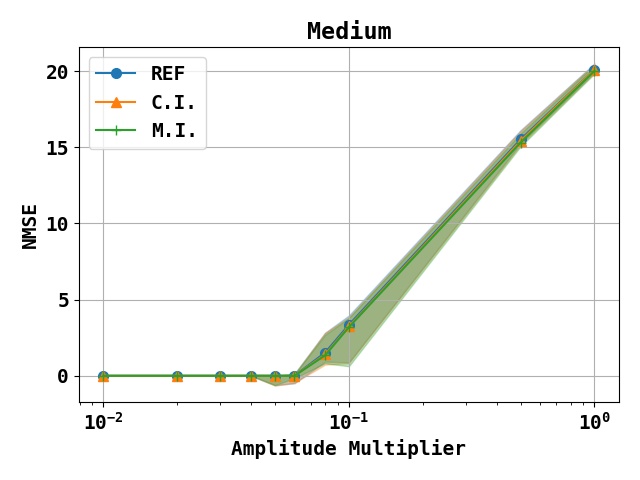}  \\
\includegraphics[height=0.2\textheight]{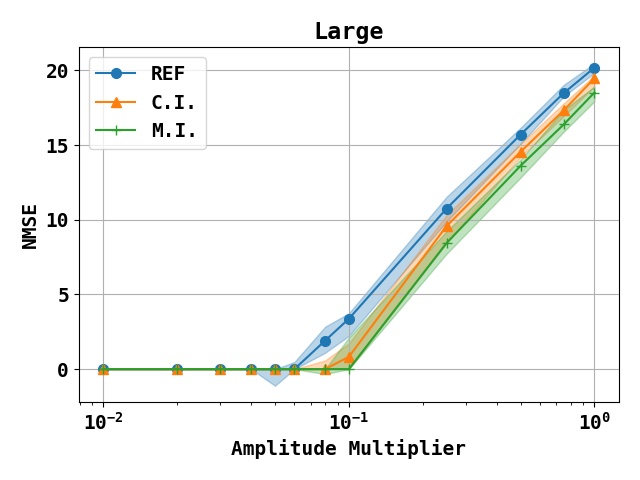} 
\end{tabular}
    \caption{Median $NMSE$ with amplitude multiplier for different types of gaps.
    \textbf{Upper row:} Small gaps. \textbf{Middle row:} Medium gaps. \textbf{Lower row:} Large gaps. 
    The final NMSE is not affected by small and medium-sized gaps. 
    However, larger gaps deteriorate the quality of extracted signal. C.I. performs better than M.I. as it virtually increases the SNR by not making up for the power loss in the gaps.
    }
    \label{fig:NMSE_with_amplitude-multiplier}
\end{figure}
The $NMSE$ evaluates the distance between the real solution $\mathcal{S}^{true}$ (the real signal without noise nor gaps) and its sparse estimate $\widehat{\mathcal{S}}$ (as the limit of the sequence $(\mathcal{S}^n)_{n\in\mathbbm{N}}$) in the time domain and is defined as follows:

\begin{equation}
\label{eq:def_MSE_global}
NMSE = -10\log_{10} \left[\frac{\displaystyle\sum_{\alpha}\norm{\mathcal{S}^{true}_\alpha - \widehat{\mathcal{S}}_\alpha}_2^2}{\displaystyle \sum_{\alpha} \norm{ \mathcal{S}^{true}_\alpha }_2^2}\right] \;.
\end{equation}
By construction the NMSE is \emph{large} when the sparse estimate $\widehat{\mathcal{S}}$ is \emph{close} to the real solution $\mathcal{S}^{true}$. 
It provides insights into the global quality of the reconstructed signal. 
However it does not provide any information about the FP and FN rates. 

Consider an input of form:
\begin{equation}
\mathcal{V}_\alpha = M(a \mathcal{S}^{true}_\alpha + \mathcal{N}^{true}_\alpha) \;,
\end{equation}
where $\mathcal{S}^{true}_\alpha$ is the signal emitted by the GB with parameters detailed in Appendix~\ref{app:GB_parameters}, received on channel $\alpha \in \{A,E\}$. 
$a \in [0,1]$ is the amplitude multiplier that we will vary during our study to simulate inputs of various amplitudes.
We examined three fixed masks, each corresponding to a different gap situation. 
For a fixed amplitude multiplier, the median NMSE is obtain as the median of the NMSEs of the solutions found for different noise realizations (for a total of 25 noise realizations ). 

In Fig.~\ref{fig:NMSE_with_amplitude-multiplier} we presented the NMSE that we obtain for an ungapped signal and for a gapped signal using the C.I. algorithm and the M.I. algorithm for the three types of gaps.
Small and medium-sized gaps have little impact on the final NMSE: the recovered signal has the same quality for a gapped and an ungapped signal. 
This is true because only a small portion of the information was lost.

With bigger gaps (corresponding to a data loss of $25\%$ in this case), the recovery results are impacted by the information loss. The amount of recovered information is still acceptable.
Nevertheless, the signal power loss is harder to compensate, which results in a deterioration of the extracted signal compared to the case without gaps.

At this stage only, we can notice a difference between the two algorithms performances: Alg.~\ref{alg:algorithm_gapped-problem_lower-frequencies} seems to perform better than Alg.~\ref{alg:all-frequencies_inpainting_algorithm}. This performance spread can be explained by the fact that Alg.~\ref{alg:all-frequencies_inpainting_algorithm} reintroduced higher frequency noise in the gaps and thus results in a noisier extracted signal.

\paragraph{Application to LDC1-3 with gaps}

We applied Alg.~\ref{alg:algorithm_gapped-problem_lower-frequencies} and Alg.~\ref{alg:all-frequencies_inpainting_algorithm} to a more realistic case. 
The dataset was produced for the LISA Data Challenge LDC1-3 which can be found online \cite{LDCwebsite}. It consists in 10 verification binaries that can be seen in Fig.~\ref{fig:information_leakage}. We modified it with the three types of gaps presented in Table~\ref{tab:gaps_parameters}, which corresponds to a loss ratio of about $27\%$, and  we evaluated the quality of the signal extracted by both algorithms. Results are presented in Table~\ref{tab:peak-to-peak_NMSE_LDC1-3} and are compared to the quality of extracted signal for an ungapped signal (computed in Ref.~\cite{blelly2020sparsity}).

\begin{table}[h!]
    \centering
    \begin{tabular}{|c|c|c|c|}
\hline
Peak & REF & C.I. & M.I. \\
\hline
1 & 9.632 & 7.566 & 8.806\\
\hline
2 & 9.172 & 7.067 & 7.936 \\
\hline
3 & 4.855 & 3.086 & 1.799\\
\hline
4 & 2.154 & 0.0  & 0.0\\
\hline
5 & 15.552 & 12.047 & 13.604\\
\hline
6 & 13.151 & 10.972 & 11.456\\
\hline
7 & 14.184 & 11.180 & 11.467\\
\hline
8 & 5.138 & 3.897 & 2.199\\
\hline
9 & 2.222 & 0.834 & 0.743\\
\hline
10 & 13.524 & 10.677 & 10.575\\
\hline
Global & 12.971 & 12.177 & 10.235\\
\hline
    \end{tabular}
    \caption{Peak to peak NMSE for  extracted signal from ungapped signal (REF), with classical inpainting (C.I.) and modified inpainting (M.I.). For this experiment the rejection rate was set to $\rho = 10^{-6}$.}
    \label{tab:peak-to-peak_NMSE_LDC1-3}
\end{table}

We computed a local NMSE corresponding to the quality of each extracted source, and a global NMSE corresponding to the quality of the total extracted signal. The NMSEs depend on the chosen algorithm and noise configuration.  The C.I. algorithm tends to better detect signals that are close to the noise level (peaks 3, 8 and 9), whereas the M.I. algorithm tends to yield better signals with high amplitude (peaks 5 6 7 and 10). On the overall, Alg.~\ref{alg:algorithm_gapped-problem_lower-frequencies} and Alg.~\ref{alg:all-frequencies_inpainting_algorithm} extract signals of similar quality but that might not be the case for other types of sources. 

\section{Conclusion}

With a foreseen duty cycle of 75\%, data gaps will constitute an important feature of realistic LISA measurements that may impact the scientific deliverables of the mission. To date only a handful of studies have addressed the problem of gap mitigation. We contribute to this effort by introducing a new non-parametric method, in the form of two complementary algorithms. Based on the sparsity framework introduced in Ref.~\cite{blelly2020sparsity} we showed that it is possible to fill the gaps with both signal and noise so that the recovered noise distribution matches the expected one and the signal power loss is compensated. 

We conducted extensive tests of this non-parametric approach and demonstrated the performances of both algorithms when confronted to different types of gaps. We also considered a more realistic case where more than $27\%$ of the data was lost in presence of multiple GB sources. The algorithms achieve similar performances in terms of noise distribution, detection capacity or accuracy of extracted signal, to situations with ungapped data. More precisely the C.I. algorithm successfully recovers the expected noise distribution but is challenged by the extraction of low SNR signals. On the contrary the M.I. algorithm yields an efficient signal reconstruction but outputs a less satisfactory noise distribution.

The current limitations of the proposed algorithms are mainly related to the signal extraction component. 
As we adopted here a model-independent approach through a representation of the signal in Fourier domain, there is natural room for improvement in the matching of the solution that we find with the expected waveforms. 
We anticipate a marked improvement of the recovered signal quality in presence of large gaps with the use of an adequate representation.
This will make the object of a future study. 

Moreover the algorithms also depend on the noise modeling through the definition of the threshold $\gamma$ with regard to noise level.
However the adaptation to the case where the noise distribution is unknown (but still supposed Gaussian in frequency domain) is straightforward and only requires an estimation of the PSD during the resolution process.

The general framework described here can be used both as a detection mean and a pre-processing step in the LISA pipeline. 
As it can be adapted to a wide range of gravitational events providing that the source admits a sparse representation on a specific domain, the present study paves the way for further investigations and extensions of this type of methods. 

\label{sec:conclusion}


\section{Acknowledgments}
The authors thank N.Korsakova and Q.Baghi and more broadly the whole LISA artifacts group for stimulating and fruitful exchanges. JB was funded by the by the European Community through the grant LENA (ERC StG - contract no. 678282).

\section{Appendix}
\appendix
\label{sec:appendix}

\section{Summary of frequently used variables}
\label{appendix:frequently_used_variables}

Table~\ref{tab:variables-summary} lists the most frequently used variables and their signification.

\renewcommand{\arraystretch}{1.5}
\begin{table*}[h!]
    \centering
    \begin{tabular}{|c|l|}
\hline
\textbf{Variable}   & \textbf{Notation} \\
\hline
True ungapped signal & $\mathcal{S}^{true}_\alpha $ \\
\hline
True ungapped noise & $\mathcal{N}^{true}_\alpha$\\ 
\hline \hline
gapped data  & $\mathcal{V},\mathcal{V}_\alpha$\\
\hline
data variable & $V, V_\alpha$\\
\hline
Data mask & $M$ \\
\hline  \hline
Fourier transform & $\widetilde{Z}$ \\
\hline
Gaussian Law & $\mathcal{G}(\mu,\sigma)$\\
\hline
PSD & $\mathbf{\Sigma}$\\
\hline  \hline
Signal variable& $S$ \\
\hline
Signal estimator & $\widehat{\mathcal{S}}$ \\
\hline
Signal sequence & $\mathcal{S}^n \to \widehat{\mathcal{S}}$ \\
\hline  \hline
Missing data variable & $U$ \\
\hline
Missing data estimator &$\widehat{\mathcal{U}}$ \\
\hline
Missing data sequence & $\mathcal{U}^n \to \widehat{\mathcal{U}}$\\
\hline  \hline
Noise variable & $N$ \\
\hline
Noise estimator & $\widehat{\mathcal{N}}$ \\
\hline
Noise sequence & $\mathcal{N}^n \to \widehat{\mathcal{N}}$ \\
\hline  \hline
Updated data & $\mathcal{V}^n$ \\
\hline
Updated noise out of gaps & $\mathcal{N}^n_{\text{gap}}=\mathcal{V}-M\mathcal{S}^n$ \\
\hline  \hline
Optimization problem for LF Eq.~(\ref{eq:BF-algo_equation})& $f_{LF}(V)$ \\
\hline
Noise sampling for HF algorithm &  $\mathcal{N}_{\text{samp}}$\\
\hline
Whitened noise estimator & $\eta = \mathbf{\Sigma}^{-1/2}\widehat{\mathcal{N}}$ \\
\hline
\end{tabular}
\caption{Summary of frequently used variables and their meaning. Calligraphic letters are used for data, estimators (denoted by a hat) and the sequential solutions of optimization problems (denoted by sequence index $n$). Capital letters are used for dummy variables.}
\label{tab:variables-summary}
\end{table*}


\section{Linearity of low frequency inpainting algorithm}
\label{app:linearity_CP-algo}

Let us denote $\Phi = \{\frac{1}{\sqrt{N_T}}\exp^{-\frac{2i\pi k n}{N_T}}\}_{0 \leq k,n \leq N_T-1}$ the matrix containing the coefficients needed to perform the discrete Fourier transform. For any measurement $V \in \mathbbm{R}^{N_T\text{x}2}$ we have :
\begin{equation}
\widetilde{V} = \Phi V \;.
\end{equation}

Let $V$ be a gapped measurement $V$ with mask $M$. We want to study the dependence of:
\begin{align}
f_{LF}(V) &= \argmin_{\substack{N \in \mathbbm{R}^{N_T\text{x}2} \\ \\ M N = V }} \frac{1}{2} \norm{\widetilde{N}}^2_{2,2,\mathbf{\Sigma}} \; \nonumber \\
&=  \argmin_{\substack{N \in \mathbbm{R}^{N_T\text{x}2} \\ \\ M N = V }} \sum_{\alpha\in \{A,E\}} \frac{1}{2} N^T_\alpha \Phi^* \mathbf{\Sigma}^{-1}_\alpha \Phi N_\alpha \;.
\end{align}
The resulting cost function is separable with respect to each channel. 
For a channel $\alpha$, the Lagrangian of this problem writes: 
\begin{equation}
\mathcal{L}_{V_\alpha}(N_\alpha,\Lambda) = \frac{1}{2} N^T_\alpha \Phi^* \mathbf{\Sigma}^{-1}_\alpha \Phi N_\alpha + \inprod{\Lambda}{V_\alpha - M N_\alpha} \;,
\end{equation}
where $\inprod{}{}$ denotes the classical hermitian inner product.

The solution of the problem can be written using the Lagrangian:
\begin{equation}
f_{LF}(V) = \bigg(\argmin_{N_\alpha} \argmax_{\Lambda} \mathcal{L}_{V_\alpha}(N_\alpha,\Lambda)\bigg)_{\alpha \in \{A,E\}} \;.
\end{equation}
The optimality conditions read: 
\begin{align}
\begin{cases}
\displaystyle \frac{\partial \mathcal{L}_{V_\alpha}}{\partial N} (N_\alpha, \Lambda) &= \Phi^* \mathbf{\Sigma}^{-1}_\alpha \Phi N_\alpha - M \Lambda = 0 \;, \\
\displaystyle \frac{\partial \mathcal{L}_{V_\alpha}}{\partial \Lambda} (N_\alpha, \Lambda) &= V_\alpha - M N_\alpha = 0 \;.
\end{cases}
\end{align}
They result in:
\begin{equation}
\begin{cases}
 N_\alpha = (\Phi^* \mathbf{\Sigma}^{-1}_\alpha \Phi)^{-1} M \Lambda \;,\\
 M N_\alpha = V_\alpha \;.
\end{cases}
\end{equation}
Thus:
\begin{equation}
\begin{cases}
 N_\alpha = (\Phi^* \mathbf{\Sigma}^{-1}_\alpha \Phi)^{-1} M \Lambda \;, \\
 M (\Phi^*\mathbf{\Sigma}^{-1}_\alpha \Phi)^{-1} M \Lambda = V_\alpha \;.
\end{cases}
\end{equation}

The solution of the Lagrangian problem thus linearly depends on $V_{\alpha}$, \ie for any measurements $V^1,V^2$ gapped with the same mask $M$, we have:
\begin{equation}
f_{LF}(V^1+V^2) = f_{LF}(V^1) + f_{LF}(V^2) \;.
\end{equation}


\section{Algorithms: proofs of convergence}

As we built Alg.~\ref{alg:algorithm_gapped-problem_lower-frequencies} as a Block Coordinate Descent (BCD) procedure (the variables $\widehat{\mathcal{S}}, \widehat{\mathcal{U}}$ minimize the cost function~(\ref{eq:cost-function_gapped-problem})), the argument of Refs.~\cite{Xu2013_BCD, Tseng:2001} applies and establishes the convergence of Alg.~\ref{alg:algorithm_gapped-problem_lower-frequencies}. 
In order to prove that Alg.~\ref{alg:all-frequencies_inpainting_algorithm} also converges, we similarly recast it as the minimization of a cost function through a BCD procedure.

\label{app:all-frequencies_convergence}

Consider a \textbf{fixed noise sample}:
\begin{equation}
    \mathcal{N}_{\text{samp}} \sim \mathcal{G}(0,\mathbf{\Sigma}) \;,
\end{equation}
corresponding to the expected noise distribution in frequency domain, and the following algorithm:
\begin{align}
\displaystyle
\begin{cases}
\mathcal{N}_{\text{gap}}^{n+1} &= \mathcal{V} - M \mathcal{S}^{n} + M \mathcal{N}_{\text{samp}} \;, \\
\mathcal{N}^{n+1}_{LF} &= \displaystyle\argmin_{\mathcal{N}_{\text{gap}}^{n+1} = M N} \frac{1}{2} \norm{\widetilde{N}}^2_{2,2,\mathbf{\Sigma}} \;, \\
\mathcal{N}^{n+1} &=\mathcal{N}^{n+1}_{LF} -  \mathcal{N}_{\text{samp}} \;, \\
\mathcal{V}^{n+1} &= \mathcal{V} + (I-M)(\mathcal{S}^{n}+\mathcal{N}^{n+1})  \;, \\
\mathcal{S}^{n+1} &= \displaystyle \argmin_{S} \norm{\gamma \odot \mathbf{\Sigma}^{-1/2} \widetilde{S}}_{1,2}
+ \norm{\widetilde{\mathcal{V}}^{n+1}-\widetilde{S}}^2_{2,2,\mathbf{\Sigma}}\;,
\end{cases}
\label{eq:app_algo_all-frequency-inpainting}
\end{align}
initialized with $\mathcal{S}^{0} = 0$.
Instead of looking for the solution as a decomposition signal/noise $(\widehat{\mathcal{S}}, \widehat{\mathcal{N}})$, we seek the decomposition in terms of signal/missing data $(\widehat{\mathcal{S}}, \widehat{\mathcal{U}})$.
To this aim, we change variables similarly to Eqs.~(\ref{eq:missing_data_equation}-\ref{eq:change_of_variable}) that link the noise variable $N$ and the missing data variable $U$. 
This relation writes: 
\begin{equation}
\mathcal{V} + M \mathcal{N}_{\text{samp}} + U = \mathcal{S}^{n} + N \;.
\label{eq:change_of_variable_alg2}
\end{equation}
By definition:
\begin{equation}
\mathcal{N}_{\text{gap}}^{n+1} = \mathcal{V} - M \mathcal{S}^{n} + M \mathcal{N}_{\text{samp}} \;,
\end{equation}
and the constraint on the noise solution reads:
\begin{equation}
\mathcal{N}_{\text{gap}}^{n+1} = M N \;,
\end{equation}
Combining both with the change of variables formula provides:
\begin{equation}
\mathcal{V} - M \mathcal{S}^{n} + M \mathcal{N}_{\text{samp}} = M \left( \mathcal{V} - \mathcal{S}^{n}+M\mathcal{N}_{\text{samp}} + U \right) \;.
\end{equation}
Thus, the constraint on the missing data variable $U$ writes:
\begin{equation}
M U = 0 \;.
\end{equation}
The equation on the missing data reads:
\begin{equation}
\mathcal{U}^{n+1}_{LF} =\displaystyle\argmin_{\substack{U \in \mathbbm{R}^{N_T\text{x}2} \\ M U = 0}} \frac{1}{2} \norm{\widetilde{\mathcal{V}}+M\widetilde{\mathcal{N}}_{\text{samp}}+\widetilde{U}-\widetilde{\mathcal{S}}^{n}}^2_{2,2,\mathbf{\Sigma}} \;,
\end{equation}
and thanks to the change of variables the following equality holds:
\begin{equation}
\mathcal{V} + M \mathcal{N}_{\text{samp}} + \mathcal{U}^{n+1}_{LF} = \mathcal{S}^{n} + \mathcal{N}^{n+1}_{LF} \;.
\end{equation}
Subtracting $\mathcal{N}_{\text{samp}}$ from both sides yields:
\begin{equation}
    \mathcal{V} + \mathcal{U}^{n+1} = \mathcal{V}^{n+1} \;,
\end{equation}

which can be plugged in the equation for $\mathcal{S}^{n+1}$:
\begin{align}
\mathcal{S}^{n+1} &= \displaystyle \argmin_{S} \norm{\gamma \odot \mathbf{\Sigma}^{-1/2} \widetilde{S}}_{1,2} \; \nonumber \\
& + \norm{\widetilde{\mathcal{V}} +  \widetilde{\mathcal{U}}^{n+1}-\widetilde{S}}^2_{2,2,\mathbf{\Sigma}} \; \nonumber \\
&= \displaystyle \argmin_{S} \norm{\gamma \odot \mathbf{\Sigma}^{-1/2} \widetilde{S}}_{1,2} \; \nonumber \\
& + \norm{\widetilde{\mathcal{V}} +  \widetilde{\mathcal{U}}^{n+1}_{LF}-(I-M)\widetilde{\mathcal{N}}_{\text{samp}}-\widetilde{S}}^2_{2,2,\mathbf{\Sigma}} \;,
\end{align}
before expanding the quadratic norm:
\begin{align}
&\norm{\widetilde{\mathcal{V}} +  \widetilde{\mathcal{U}}^{n+1}_{LF}-(I-M)\widetilde{\mathcal{N}}_{\text{samp}}-\widetilde{S}}^2_{2,2,\mathbf{\Sigma}} = \; \nonumber \\
&\norm{\widetilde{\mathcal{V}} +  \widetilde{\mathcal{U}}^{n+1}_{LF}+M\widetilde{\mathcal{N}}_{\text{samp}}-\widetilde{S}}^2_{2,2,\mathbf{\Sigma}} \; \nonumber \\ 
&+ \norm{\widetilde{\mathcal{N}}_{\text{samp}}}^2_{2,2,\mathbf{\Sigma}} \; \nonumber \\
&- 2\Re \inprod{\widetilde{\mathcal{V}} +  \widetilde{\mathcal{U}}^{n+1}_{LF}+M\widetilde{\mathcal{N}}_{\text{samp}}-\widetilde{S}}{\widetilde{\mathcal{N}}_{\text{samp}}} \;.
\end{align}
As we are optimizing with regard to the variable $S$, we can remove all the terms that are independent of it. Finally $\mathcal{S}^{n+1}$ is defined as:
\begin{align}
\mathcal{S}^{n+1} = \displaystyle &\argmin_{S} \norm{\gamma \odot \mathbf{\Sigma}^{-1/2} \widetilde{S}}_{1,2} \; \nonumber \\
&+ \frac{1}{2} \norm{\widetilde{\mathcal{V}} +  \widetilde{\mathcal{U}}^{n+1}_{LF}+M\widetilde{\mathcal{N}}_{\text{samp}}-\widetilde{S}}^2_{2,2,\mathbf{\Sigma}} \; \nonumber \\
&+ \Re \inprod{\widetilde{S}}{\widetilde{\mathcal{N}}_{\text{samp}}} \;.
\end{align}

The cost function:
\begin{align}
J_{\text{M.I.}}&(S,U) = \displaystyle \norm{\gamma \odot \mathbf{\Sigma}^{-1/2} \widetilde{S}}_{1,2} \; \nonumber \\
&+ \frac{1}{2} \norm{\widetilde{\mathcal{V}} +  \widetilde{U}+M\widetilde{\mathcal{N}}_{\text{samp}}-\widetilde{S}}^2_{2,2,\mathbf{\Sigma}} \; \nonumber  \\
&+ \Re \inprod{\widetilde{S}}{\widetilde{\mathcal{N}}_{\text{samp}}} \;,
\end{align}
is block-convex and its minimum can be reached through: 
\begin{align}
\begin{cases}
\mathcal{U}^{n+1}_{LF} &= \displaystyle\argmin_{\substack{U \in \mathbbm{R}^{N_T\text{x}2} \\ M U = 0}}  J_{\text{M.I.}}(\mathcal{S}^n,U) \;, \\
\mathcal{S}^{n+1} &= \displaystyle\argmin_{S} J_{\text{M.I.}}(S,\mathcal{U}^{n+1}_{LF}) \;, \\
\end{cases}
\end{align}
with initialization $\mathcal{S}^0 = 0$. We recognize the form of a BCD algorithm, hence justifying the convergence of the system~(\ref{eq:app_algo_all-frequency-inpainting}) to the global minimum of the cost function $J_{M.I.}$.


\section{Solving the noise with the Chambolle-Pock algorithm}
\label{app:noise_solver_CP-algorithm}

To keep notations compact, we will respectively denote $D = \mathbbm{R}^{N_T \times 2}$ and $\mathcal{D} = \mathbbm{C}^{(2N_f+1) \times 2}$ the time and frequency domains. 

\subsection{Primal-dual formulation}

Chambolle and Pock developed the primal-dual algorithm to solve problems with the general form:
\begin{equation}
\label{eq:chambolle-pock-primal}
\argmin_{x \in D} G(x) + F(Kx) \;,
\end{equation} 
where $K$ is a matrix, $F$ and $G$ are convex functions also satisfying some extra assumptions \cite{chambolle:hal-00490826} not reminded here.

Denoting by $K^*$ the conjugate operator of $K$ and by $F^*, G^*$ the conjugate applications of $F, G$ (see Ref.~\cite{boyd_convex_2004} for a definition of conjugate applications), it was shown in \cite{chambolle:hal-00490826} that using the following primal-dual algorithm:
\begin{align}
\label{eq:chambolle-pock-iteration}
\begin{cases}
y^{p+1} &= \prox_{\lambda F^*} \left(y^p + \lambda K \overline{x}^p \right) \;, \\
x^{p+1} &= \prox_{\tau G} \left( x^p - \tau K^* y^{p+1} \right) \;, \\
\overline{x}^{p+1} &= x^{p+1} + \theta (x^{p+1}-x^p) \;.\\
\end{cases}
\end{align}
the sequence $(x^p)_{p \in \mathbbm{N}}$ converges to the solution of the optimization problem Eq.~(\ref{eq:chambolle-pock-primal}). 
The induction (\ref{eq:chambolle-pock-iteration}) can be initialized with arbitrary $x^0 \in D, y^0 \in \mathcal{D}$ and $\overline{x}^0 = x^0$. 
The parameters $\theta \in (0,1)$ and $\lambda,\tau$ are chosen to fulfill the criterion:
\begin{equation}
\tau \lambda L^2 < 1 \;,
\label{eq:CP_parameters_constraint}
\end{equation}
where $L$ denotes the norm of the matrix $K$. The proximal function $\prox_{\alpha f}$ is defined for a function $f$ and a real $\alpha > 0$ by:
\begin{equation}
\prox_{\alpha f}(u) = \argmin_z \left[ \alpha f(z) + \frac{1}{2} \norm{z-u}^2_{2,2} \right] \;.
\end{equation}
At last the matrix $K^*$ is the adjoint of $K$.

In the present context of gapped data in LISA, we have to minimize a function defined in the Fourier domain $\mathcal{D}$ (where the PSD is diagonal) subjected to the constraint $U \in \mathrm{Ker}(M)$ expressed in the time domain $D$. 
The inpainted noise within the gaps is solution of Eq.~(\ref{eq:classical-inpainting_noise-update}) reminded here:
\begin{align}
\mathcal{N}^{n+1}_{LF} &= \displaystyle\argmin_{\mathcal{N}^{n+1}_{gap} = M N} \frac{1}{2} \norm{\widetilde{N}}^2_{2,2,\mathbf{\Sigma}} \nonumber \\
&= \displaystyle\argmin_{N} \mathbbm{1}\left[ \mathcal{N}^{n+1}_{gap} = M N \right] + \frac{1}{2} \norm{\widetilde{N}}^2_{2,2,\mathbf{\Sigma}} \;,
\label{eq:CP_noise-equation}
\end{align}
where $\mathcal{N}^{n+1}_{gap} = \mathcal{V} - \mathcal{S}^{n}$ and the characteristic function $\mathbbm{1}$ satisfies:
\begin{equation}
   \mathbbm{1}[\mathcal{N}^{n+1}_{gap} = M N] = \begin{cases} +\infty \text{ if } \mathcal{N}^{n+1}_{gap} = M N \\ 0 \text{ otherwise} \end{cases} \;.
\end{equation}

This problem is amenable to a resolution with the Chambolle-Pock algorithm with the following identification: 

\begin{align}
\begin{cases}
KV &= \mathbf{\Sigma}^{-1/2} \widetilde{V} \;, \\
F(\widetilde{V}) &= \frac{1}{2}\norm{\widetilde{V}}^2_{2,2} \;, \\
G_n(V) &= \mathbbm{1}[\mathcal{V} - M\mathcal{S}^{n} = M V] \;.
\end{cases}
\end{align} 
After some algebra involving Moreau's identity, we obtain:
\begin{align}
\prox_{\tau G_n} (V) &= (U - M S^{n+1}) + (I-M) V \;,\\
\prox_{\alpha F}(\widetilde{V}) &= \frac{1}{1+\alpha} \widetilde{V} \;, \\
\prox_{\lambda F^*}(V) &= \frac{1}{1+\lambda} V \;. 
\end{align}

\subsection{Preconditioned formulation}

The parameters $\tau$ and $\sigma$ are constrained by Eq.~(\ref{eq:CP_parameters_constraint}) involving the norm of the matrix $K$, \ie the norm of the inverse square root of the noise PSD. Since this PSD is very ill-conditioned, we will have to select very small $\tau, \lambda$ to satisfy Eq.~(\ref{eq:CP_parameters_constraint}) which leads to a slow convergence. This calls for a preconditioning of the primal-dual formulation along the lines of Ref
~\cite{pock_CPdiag}. The main idea consists in changing the inner products equipping the time $D$ and frequency $\mathcal{D}$ domains. This results in a mere modification of the proximal operators used in the iteration (\ref{eq:chambolle-pock-iteration}) which leaves its computational complexity basically unchanged. 

We thus define new $\prox$ operators using two symmetric definite positive matrices $T$ and $\Lambda$: \begin{align}
\prox^D_G(u) &= \argmin_{z} G(z) + \frac{1}{2}\norm{u-z}^2_{2,2,T} \\
\prox^\mathcal{D}_F(u) &= \argmin_{z} F(z) + \frac{1}{2}\norm{u-z}^2_{2,2,\Lambda}
\end{align}
The preconditioned primal-dual algorithm writes:
\begin{align}
\begin{cases}
y^{p+1} &= \prox^D_G \left( y^p + \Lambda K \overline{x}^p \right) \;, \\
x^{p+1} &= \prox^\mathcal{D}_F \left( x^p - TK^* y^{p+1} \right) \;, \\
\overline{x}^{p+1} &= x^{p+1} + \theta (x^{p+1} - x^p) \;,
\end{cases}
\end{align}
with the same initialization as before at arbitrary $x^0 \in D, y^0 \in \mathcal{D}$ and $\overline{x}^0 = x^0$.
This algorithm converges if the norm of the matrix $\Lambda^{1/2} K T^{1/2}$ is (strictly) smaller than 1.

In the case of LISA data with a known PSD $\Sigma$, we chose the following:
\begin{align}
\Lambda &= \mathbf{\Sigma}^{1/2} \;, \\
T &= \min(\mathbf{\Sigma}^{1/2}) \;.
\end{align}
Thanks to this choice, the $\prox$ operators can still be computed in closed form.


\section{Galactic binary parameters}
\label{app:GB_parameters}

The study was conducted choosing the following parameters for the considered GBs:
\begin{description}
\item[Frequency $f_0$] = 3 mHz                       
\item[Frequency derivative $\dot f_0$] = $2.04973995 \cdot 10^{-18} \textrm{Hz}^2$ 
\item[Ecliptic latitude $\beta$] = 0. Rad       
\item[Ecliptic longitude $\lambda$] = -2.18009 Rad      
\item[Amplitude $A$] $= 1.76276 \cdot 10^{-22}$ strain       
\item[Inclination $\iota$] = 0.523599 Rad            
\item[Polarization $\psi$] = 3.61909315 Rad          
\item[Initial phase $\phi_0$] = 2.97459105 Rad          
\end{description}

These parameters are the one needed to create a GB signal by the MLDC code Ref.~\cite{LDCwebsite}.

\section{Algorithms Parameters}

\begin{description}
    \item[Signal extraction:] We refer to Ref.~\cite{blelly2020sparsity} for details about the implementation of the signal extraction algorithm for ungapped data and the tuning of its parameters.
    \item[$\epsilon = 10^{-3}$:] Convergence parameter for the BCD algorithm (global convergence parameter).
    \item[$\epsilon_{CP} = 10^{-5}$:] Convergence parameter for the Chambolle-Pock algorithm.
    \item[$N_{\text{it,BCD}}=20$:] Maximal number of iterations for the BCD algorithm.
    \item[$N_{\text{it,CP}}=500-2000$:] Maximal number of iterations for the Chambolle-Pock algorithms for small/medium and large gaps.
\end{description}


\section{Open source code}
\label{appendix:opensource_code}

The code is open source and can be found online at \url{https://github.com/GW-IRFU/gw-irfu} on version 3 of the GPL (GPLv3).

\bibliography{biblio.bib} 
\bibliographystyle{unsrt}

\end{document}